\documentclass[lettersize,journal]{IEEEtran}
\usepackage{amsmath,amsfonts}
\usepackage{algorithmic}
\usepackage{algorithm}
\usepackage{array}
\usepackage[caption=false,font=normalsize,labelfont=sf,textfont=sf]{subfig}
\usepackage{textcomp}
\usepackage{stfloats}
\usepackage{url}
\usepackage{verbatim}
\usepackage{graphicx}
\usepackage{cite}

\usepackage{amsmath}
\usepackage{graphicx}
\usepackage{subfig}
\usepackage{booktabs}
\usepackage{threeparttable}
\usepackage{hyperref}
\usepackage{utfsym}
\usepackage{multirow}
\usepackage{makecell}
\usepackage{tikz,xcolor}
\hypersetup{hidelinks,
	colorlinks=true,
	allcolors=black,
	pdfstartview=Fit,
	breaklinks=true}
 \definecolor{lime}{HTML}{A6CE39}
\DeclareRobustCommand{\orcidicon}{
\begin{tikzpicture}
\draw[lime, fill=lime] (0,0)
circle[radius=0.16]
node[white]{{\fontfamily{qag}\selectfont \tiny \.{I}D}};
\end{tikzpicture}
\hspace{-2mm}
}
\foreach \x in {A, ..., Z}{%
\expandafter\xdef\csname orcid\x\endcsname{\noexpand\href{https://orcid.org/\csname orcidauthor\x\endcsname}{\noexpand\orcidicon}}
}

\begin{document}

\title{Ultra-Low Complexity On-Orbit Compression for Remote Sensing Imagery via Block Modulated Imaging}
 
\author{
Zhibin~Wang\hspace{-1.5mm}\orcidA{}, 
Yanxin~Cai\hspace{-1.5mm}\orcidG{}, 
Jiayi~Zhou\hspace{-1.5mm}\orcidE{}, 
Yangming~Zhang\hspace{-1.5mm}\orcidD{}, 
Tianyu~Li*\hspace{-1.5mm}\orcidF{}, 
Wei~Li\hspace{-1.5mm}\orcidH{}, 
Xun~Liu\hspace{-1.5mm}\orcidI{}, 
Guoqing~Wang*\hspace{-1.5mm}\orcidC{},~\IEEEmembership{Member,~IEEE, }
and~Yang~Yang\hspace{-1.5mm}\orcidB{},~\IEEEmembership{Senior~Member,~IEEE}
\thanks{Manuscript created August, 2024; This work was supported in part by the National Natural Science Foundation of China under grant U23B2011, 62102069, U20B2063 and 62220106008 \emph{(Corresponding author: Guoqing Wang and Tianyu Li.)}.
Zhibin Wang, Jiayi Zhou, Yangming Zhang, Tianyu Li, Guoqing Wang, and Yang Yang are with the School of Computer Science and Engineering, University of Electronic of Science and Technology of China, Chengdu, China (e-mail: 202321080335@std.uestc.edu.cn; zjy1551965517@gmail.com; 1272713164@qq.com; cosmos.yu@hotmail.com; gqwang0420@uestc.edu.cn; yang.yang@uestc.edu.cn).
Yanxin Cai, Wei Li and Xun Liu are with Beijing Institute of Space Mechanics and Electricity, Beijing, China (e-mail: yx\_cyx@126.com; wei\_li\_bj@163.com; liuxun\_laby@163.com).}

}

\markboth{Journal of \LaTeX\ Class Files,~Vol.~14, No.~8, August~2021}%
{Shell \MakeLowercase{\textit{et al.}}: A Sample Article Using IEEEtran.cls for IEEE Journals}


\maketitle

\begin{abstract}
The growing field of remote sensing faces a challenge: the ever-increasing size and volume of imagery data are exceeding the storage and transmission capabilities of satellite platforms. Efficient compression of remote sensing imagery is a critical solution to alleviate these burdens on satellites. However, existing compression methods are often too computationally expensive for satellites. With the continued advancement of compressed sensing theory, single-pixel imaging emerges as a powerful tool that brings new possibilities for on-orbit image compression. However, it still suffers from prolonged imaging times and the inability to perform high-resolution imaging, hindering its practical application. This paper advances the study of compressed sensing in remote sensing image compression, proposing Block Modulated Imaging (BMI). By requiring only a single exposure, BMI significantly enhances imaging acquisition speeds. Additionally, BMI obviates the need for digital micromirror devices and surpasses limitations in image resolution. Furthermore, we propose a novel decoding network specifically designed to reconstruct images compressed under the BMI framework. Leveraging the gated 3D convolutions and promoting efficient information flow across stages through a Two-Way Cross-Attention module, our decoding network exhibits demonstrably superior reconstruction performance. Extensive experiments conducted on multiple renowned remote sensing datasets unequivocally demonstrate the efficacy of our proposed method. To further validate its practical applicability, we developed and tested a prototype of the BMI-based camera, which has shown promising potential for on-orbit image compression. The code is available at \url{https://github.com/Johnathan218/BMNet}.
\end{abstract}


\section{Introduction}

\begin{figure*}[ht]
    \centering
    \includegraphics[width=\textwidth]{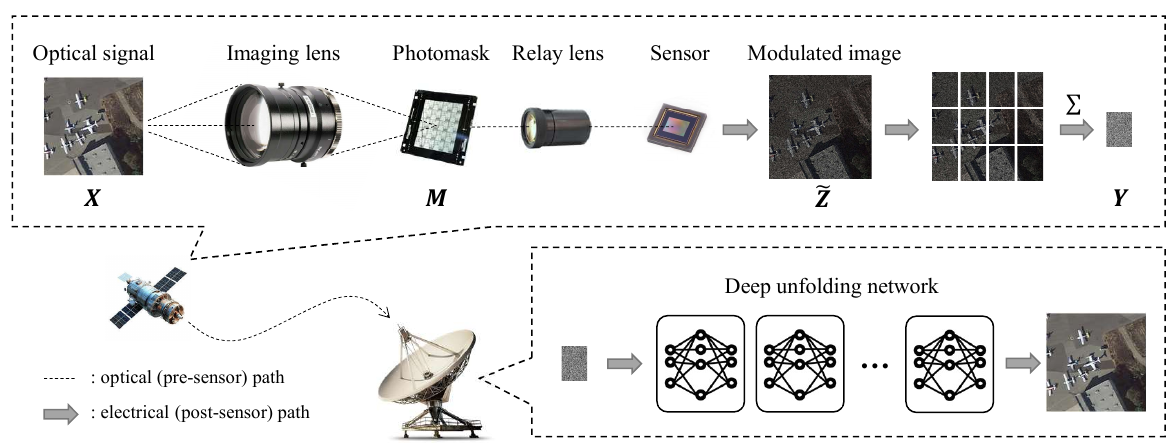}
    \caption{Satellite-side optical encoding for compression and ground-based deep unfolding decoding for reconstruction.}
    \label{encodingscheme}
\end{figure*}

\IEEEPARstart{O}{ptical} remote sensing images are instrumental across various fields, providing high-resolution data that informs decision-making, optimizes resource allocation, and mitigates disaster risks \cite{10239453, 8350339, 4215094}. Recent advancements in remote sensing have led to a substantial increase in both the spatial resolution of images and the volume of data collected \cite{7565634,10439079}. 
The acquisition of high-resolution imagery by remote sensing satellites generates vast amounts of data that strain onboard storage capacity, data transmission bandwidth, and energy resources. Consequently, onboard processing, storage, and communication capabilities are all significantly impacted, hindering real-time  processing and satellite-to-ground/inter-satellite data transmission. 

Remote sensing image compression algorithms are crucial techniques for addressing the storage and transmission challenges posed by massive remote sensing image collections \cite{9144368}. Onboard image compression algorithms fall into two main categories \cite{RN52}: lossless and lossy. Lossless compression exploits redundancies within the image data to achieve smaller file sizes, while maintaining a perfect reconstruction of the original image without any information loss. However, lossless compression typically achieves low compression ratios \cite{HUSSAIN201844}, making it inadequate for handling the transmission bandwidth and storage limitations of satellites. In contrast, lossy compression offers significantly higher compression ratios by tolerating a controlled degree of information loss. However, this advantage comes at the cost of increased computational demands, which can strain the limited processing power available onboard spacecraft. The increased computational burden imposed by complex compression algorithms can lead to performance degradation of the satellite's processor, potentially impacting the execution of other critical tasks. 
Extensive research \cite{6626242, 7026133, 2017.0469, malathkar2020near} has been dedicated to addressing the challenge of constrained computational resources on specific platforms. Numerous strategies have been advanced to reduce the encoding complexity of compression algorithms. Nevertheless, these methodologies have not yielded significant alterations, and the burden on satellite processors remains considerable for high-resolution imaging.

Building upon the pioneering work of Compressed Sensing \cite{1614066} (CS), which exploits signal sparsity for sub-Nyquist sampling \cite{1697831}, Single Pixel Imaging \cite{4472247, 4752747, 52362346234, adfdsadfb, 8588994} (SPI)  unveils an entirely novel scheme and introduces enhanced alternatives for the compression of remote sensing imagery. In contrast to conventional compression methods that perform computations in the electrical domain after capturing the images using sensors, SPI performs the computation in the optical domain through the optics prior to sensor detection, directly capturing a compressed representation of the scene.
Compared to traditional compression methods like JPEG \cite{10.1145/103085.103089} and JPEG2000 \cite{952804}, SPI features a far less complex encoding process, making it well-suited for resource-constrained platforms. However, the application of SPI in satellite scenarios encounters two substantial obstacles. The first challenge is that Single Pixel Cameras (SPCs) can only capture one measurement at a time yet acquiring high-resolution images typically necessitates tens of thousands of measurements \cite{elkdhssdf, Monin2021}. It necessitates prolonged acquisition periods for time-series measurements, a factor that requires static imaging and is particularly unfavorable for satellites in motion. The limited imaging speed can also considerably impact the execution of real-time missions. The second challenge is that SPCs rely on Digital Micromirror Devices (DMDs) for dynamic encoding. Unfortunately, the limited size and modulation frequency of DMDs cannot accommodate the high-resolution requirements of remote sensing imagery \cite{Li:21, Stojek:22}. It is also worth mentioning that current SPI decoding algorithms exhibit high computational complexity and limited suitability for high-resolution images, as evidenced in our experiments in Section \ref{highre}.

In this paper, we propose a novel compression paradigm for remote sensing images, termed Block Modulated Imaging (BMI), which not only inherits the benefits of CS with ultra-low encoding complexity in comparison to conventional methods, but also exhibits superior practicality when contrasted with SPI. Inspired by BMVC \cite{xue2022block}, our encoding scheme involves a simple modulation of the image signal utilizing a photomask, followed by slicing and summation operations applied to the modulated image, as shown in Fig. \ref{encodingscheme}. The whole process has a negligible impact on computing resources. Superior to SPI, our method requires only a single exposure, thereby eliminating the need for static imaging. Furthermore, the necessity for DMDs has been obviated. We can now utilize a pre-configured optical mask, which is capable of adapting to high-resolution imagery.
Based on the theory of CS \cite{Yuan_2020_CVPR}, the image can be restored with high quality by designing a suitable deep neural network for decoding. We introduce a novel decoding network based on the deep unfolding architecture \cite{meng2020gap}. Specifically, our model leverages gated 3D convolutions to achieve superior recovery performance, while concurrently employing cross-attention mechanisms to facilitate robust information exchange across all stages. Experiments show that BMI demonstrates competitive, if not superior, decoding performance to SPI. At the same time, our decoding algorithm is less computationally intensive compared to SPI and is more suitable for high-resolution images. Furthermore, to bridge the gap between theory and practice, we presents a functional hardware prototype of the our compression system. Real-world testing of the prototype validated the feasibility of both the encoding scheme and the reconstruction algorithm.

The contributions of this paper are summarized as follows:
\begin{enumerate}
\item We propose a novel CS-based compression scheme for remote sensing imagery, termed as BMI, characterized by significantly reduced encoding complexity compared to traditional methods and superior practicality compared to SPI.
\item We develope a robust deep unfolding network for decoding, which, in comparison to SPI paradigms, has yielded improved performance. Concurrently, we establish a baseline for downstream tasks and investigate the influence of varying compression ratios on its performance.
\item A hardware prototype of our approach has been developed, demonstrating the feasibility of our compression system and paving the way for its application in real-world scenarios.
\end{enumerate}

\section{Related Work}
\subsection{Encoding Complexity Optimization for Remote Sensing Image Compression}
High-resolution optical satellites typically employ various compression techniques for remote sensing image data \cite{7153923}, including Differential Pulse Code Modulation (DPCM), Adaptive DPCM (ADPCM), Discrete Cosine Transform (DCT), and Discrete Wavelet Transform (DWT). Compared to DPCM and ADPCM, DCT and DWT offer superior compression ratios. Consequently, DCT and DWT have become more prevalent compression techniques, exemplified by the widely used JPEG \cite{10.1145/103085.103089} and JPEG2000 \cite{952804} algorithms. However, their significant demands on processing power may strain satellite platforms with limited computational resources. In recent years, reducing encoding complexity has become a focal point of research in onboard image compression algorithms. 
To address the high complexity of JPEG2000, X. Chen et al. \cite{6626242} proposed an efficient and low-complexity rate control algorithm to reduce memory requirements and computation time. 
M. Conoscenti et al. \cite{7570236} improved predictive lossy compression schemes, proposing a simple and effective rate control algorithm that significantly improves performance and reduces complexity. 
T. Ma \cite{2017.0469} proposed an efficient quadtree search model to replace the arithmetic coding in the SPIHT algorithm reducing complexity. 
J. Bartrina-Rapesta et al. \cite{7935537} designed a lightweight on-board entropy encoder, the proposed arithmetic encoder uses only bitwise operations to estimate the correlation probabilities and consumes little computational resources.
Although these methods have achieved some success in reducing encoding complexity, they still place a substantial burden on the digital processor, which is critical in resource-limited on-orbit situations.

\subsection{Single Pixel Imaging in Remote Sensing}
CS has emerged as a powerful paradigm in image compression \cite{13-6001-5_42}, offering superior compression efficiency and higher achievable compression ratios by circumventing the limitations of the Nyquist-Shannon sampling theorem \cite{1697831}. One key application of this technique is Single Pixel Imaging (SPI), which was first proposed in \cite{4472247}. 
In contrast to traditional methods that capture the entire image and then compress it computationally, SPI directly acquires a compressed representation of the scene during the sampling process. This eliminates the need for a separate compression step and leads to significant computational savings.
J. Ma \cite{4752747} pioneered the application of SPI to remote sensing, leading to improvements in power consumption, data storage, and transmission on satellites. 
Building on the principle of \cite{4752747}, J. Ma \cite{52362346234} further designed a SPI system for line-scan onboard cameras. 
The motion blur caused by the relative movement between remote sensing platforms and the target scene has been a subject of investigation by Z. Wang et al. \cite{adfdsadfb}. 
J. Li et al. \cite{8588994} leveraged DWT to maximize the use of textural features in remote sensing images for guiding sensing resource allocation. 
J. Bobak et al. \cite{8898608} discussed the benefits of adopting a single pixel camera-like architecture for microwave radiometry in terms of reducing size, weight, power consumption, and cost.
While the application of the CS in remote sensing has received increased interest, limitations of SPI in terms of imaging speed and spatial resolution have severely restricted its practical implementation. 

\begin{figure*}[t]
    \centering
    \includegraphics[width=\textwidth]{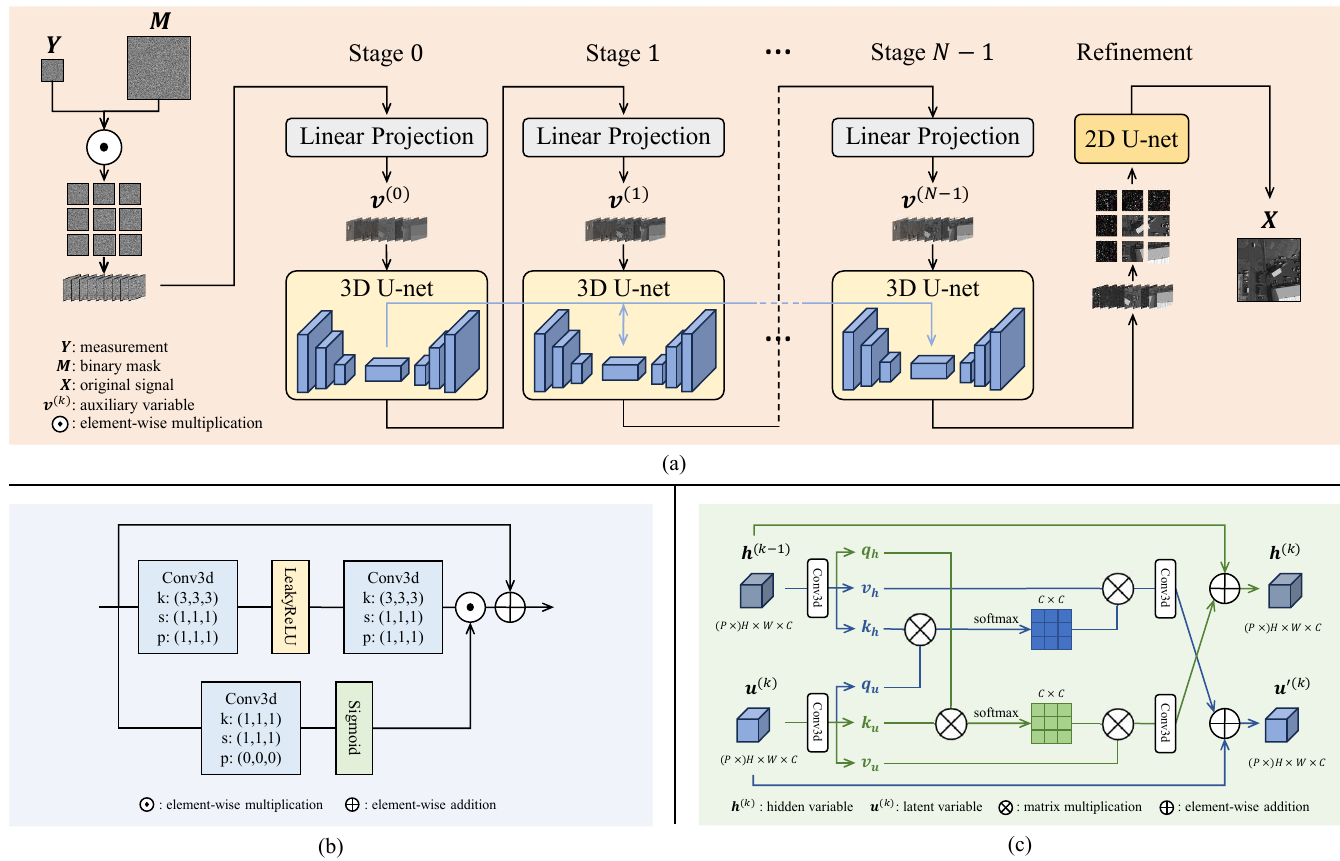}
    \caption{(a) BMNet processes the measurement $Y$ alongside the mask $M$, and reconstructs the original signal $X$. $\textbf{\emph{v}}^{(k)}$ denotes the auxiliary variable at the $k$-th stage (see Eq. \ref{linearpj}). (b) The proposed gated 3D convolution block. (c) The proposed Two-Way Cross-Attention module.  $\textbf{\emph{u}}^{(k)}$ denotes the latent variable generated by the encoder of the $k$-th 3D U-net. $\textbf{\emph{h}}_{h}^{(k)}$ denotes the hidden variable of the $k$-th stage ($\textbf{\emph{h}}^{(0)}=\textbf{\emph{u}}^{(0)}$).}
    \label{framework}
    \vspace{-1em}
\end{figure*}

\section{Method}
\subsection{Block Modulated Encoding}
\label{encoding}
In this section we present the mathematical model for the encoding process of the proposed block modulated imaging. 

We employ a grayscale image for illustration purposes. Fig. \ref{encodingscheme} demonstrates the image encoding process. Given an optical signal $X\in \mathbb{R}^{H\times W}$, after passing through the imaging lens, $X$ is modulated by a photomask. 
This modulation process can be viewed as a Hardman product:
\begin{equation}
    \label{modulation}
    Z=M\odot X
\end{equation}
where $Z\in \mathbb{R}^{H\times W}$ is the modulated optical signal and $M\in \mathbb{R}^{H\times W}$ represents a binary mask, wherein the value 1 denotes translucency and 0 indicates opacity. The 0-1 mask can be viewed as a sparse representation, aligning with the principles of compressed sensing. This mathematical framework allows for the reconstruction of the signal from a smaller number of measurements.

The modulated optical signal $Z$ is transformed into electrical data $\widetilde Z$ upon being imaged by the sensor. Afterwards, the image $\widetilde Z$ is partitioned into $N$ pieces of equal size and subsequently summed into one piece:
\begin{equation}
    \label{summation}
    Y=\sum^{N}_{i=1} \widetilde Z_{i}
\end{equation}
where $\widetilde Z_{i}$ is the $i$-th piece of the image $\widetilde Z$.

Through this, we obtain the signal $Y$ that is to be transmitted. We now turn our attention to the encoding complexity of the BMI. The Hardman product (Eq. (\ref{modulation})) is operated in the optical domain by the optics. As light traverses through the photomask, computations are executed instantaneously. The entire process is isolated from the satellite's other systems, thereby not requiring any allocation of processor resources.
The chunking and summation of the modulated image, following the conversion to electrical data, constitute the primary sources of computational complexity within the encoding process. The theory of compressed sensing enables this straightforward encoding method, offering significant computational savings compared to traditional compression algorithms like JPEG and JPEG2000. Compressed sensing fundamentally shifts the computational burden from encoding to decoding, making it particularly advantageous in applications where encoding and decoding processes can be separated, such as satellite imaging.

In contrast to the limitations of the SPI paradigm \cite{4472247}, BMI necessitates merely a single exposure to acquire the compressed representation of the image. It not only accelerates image acquisition, making it well-suited for real-time tasks, but also mitigates the issue of blurred images caused by satellite motion.
Furthermore, it enables BMI to eliminate the need for DMDs and utilize a pre-set photomask to achieve modulation, addressing the limitation of commercially available DMDs in handling high-resolution imagery. Additionally, omitting DMDs deployment could result in significant energy conservation.

\subsection{Deep Unfolding Decoding}
At the receiver end, we reconstruct the image utilizing a deep unfolding algorithm \cite{NIPS2016_1679091c,Ma_2019_ICCV,meng2020gap}. BMI and video Snapshot Compressive Imaging (SCI) \cite{9363502,9711280,10377358,bi,Cheng_2021_CVPR} share the same decoding principles. The decoding of BMI can be regarded as a more challenging task compared to the decoding of video SCI, as the correlations between blocks in images are not necessarily guaranteed.

Let $\textbf{\emph x}=[\textbf{\emph x}_1^T,\ldots,\textbf{\emph x}_B^T]^T$, $\mathbf{\Phi}=[D_1,\ldots,D_N]^T$ where $\textbf{\emph x}_i={\rm vec}(X_i)$ and $D_i={\rm diag(vec}(M_i))$, $M_i$ is the $i$-th piece of the mask $M$ corresponding to $\widetilde Z_i$. Let $\textbf{\emph{y}}={\rm vec}(Y)$, the vector representation of Eq. (\ref{summation}) is:
\begin{equation}
    \textbf{\emph{y}}=\mathbf{\Phi}\textbf{\emph x}
\end{equation}

To reconstruct the image signal $\textbf{\emph{x}}$ from measurement $\textbf{\emph{y}}$ is to solve an ill-posed optimization problem:
\begin{equation}
    \label{ill}
    \textbf{\emph{x}}=\mathop{\arg\min}\limits_{\textbf{\emph{x}}}\frac{1}{2}||\textbf{\emph{y}}-\mathbf{\Phi}\textbf{\emph{x}}||^2_2+\lambda R(\textbf{\emph{x}})
\end{equation}
where $\lambda R(\textbf{\emph{x}})$ is the prior regularization.

Here we utilize the deep unfolding algorithm to solve Eq. (\ref{ill}). The fundamental concept of deep unfolding algorithms involves unrolling the iterative steps of conventional optimization algorithms like ADMM \cite{MAL-016} and GAP \cite{gap} as phases within a deep neural network. Each phase contains a linear projection and a small neural network. Starting from $\textbf{\emph{x}}^{(0)}$, the $k$-th iteration of solving for $\textbf{\emph{x}}^{(k)}$ is: 
\begin{equation}
    \label{linearpj}
    \textbf{\emph{v}}^{(k)}=\textbf{\emph{x}}^{(k-1)}+\mathbf{\Phi}^T (\mathbf{\Phi \Phi}^T +\eta^{(k)})^{-1} (\textbf{\emph{y}}-\mathbf{\Phi}\textbf{\emph{x}}^{(k-1)})
\end{equation}
\begin{equation}
    \label{deepnn}
    \textbf{\emph{x}}^{(k)} = \mathcal{D}^{(k)}(\textbf{\emph{v}}^{(k)})
\end{equation}
where $\textbf{\emph{v}}^{(k)}$ is the auxiliary variable introduced to reconstruct $\textbf{\emph{x}}^{(k)}$, $\mathcal{D}^{(k)}$ corresponds to the neural network for the $k$-th iteration. Inspired by \cite{9711280}, we add a regularizer $\eta^{(k)}$ for each stage.
The comprehensive structure of the deep unfolding network for decoding is depicted in Fig. \ref{framework} (a), termed BMNet. BMNet contains $N$ stages along with a refinement phase. Each individual stage is composed of a linear projection and a small neural network. The linear projection at the $k$-th stage is represented by Eq. (\ref{linearpj}), while the subsequent neural network is depicted by Eq. (\ref{deepnn}).

Within the domain of image restoration, conventional approaches often involve designing neural networks based on 2D convolutional operations. However, such methods exhibit inherent limitations in modeling inter-block relationships, which is a critical aspect of the decoding process addressed in this study. Specifically, during image encoding and compression, spatial positional coherence between adjacent blocks is lost, a factor that must be explicitly accounted for during reconstruction. To overcome these limitations, we propose the integration of 3D convolutional layers into the network architecture. By stacking image blocks $\widetilde Z_{i}$ to transform a 2D image into a 3D data cube (as shown in Fig. \ref{framework} (a)), 3D convolution enables joint learning of spatial and structural dependencies across feature hierarchies, thereby addressing the challenges posed by disrupted positional information during encoding. In this paper, each stage incorporates a 3D U-net architecture. 
To further enhance the representation of BMNet, we introduce a gating mechanism \cite{Yu_2019_ICCV} for the 3D convolution, as shown in Fig. \ref{framework} (b). The initialized measurement for recovery contains invalid pixels. Due to this, applying a convolution kernel with equal weights to all pixels may not be optimal. By introducing an additional branch to compute weights, the network can dynamically select important features. 


Within deep unfolding algorithms, information flow and integration between stages are crucial for achieving optimal reconstruction performance \cite{9711280,Song_2023_CVPR,10.1145/3474085.3475562}. Optimizing these information exchange mechanisms can mitigate information loss inherent to the unfolding process. To address this challenge, we propose a novel Two-Way Cross-Attention (TWCA) module to achieve sufficient inter-stage information interaction by fully exploiting the latent vectors generated by the 3D U-net encoder, as shown in Fig. \ref{framework} (a) and (c). Let ${\rm Enc}^{(k)}(\cdot)$ and ${\rm Dec}^{(k)}(\cdot)$ denote the encoder and the decoder of the 3D U-net in $\mathcal{D}^{(k)}$ respectively. Let $\textbf{\emph{u}}^{(k)}\in \mathbb{R}^{C\times P\times H \times W}$ represents the latent variable generated by ${\rm Enc}^{(k)}(\cdot)$. We have:
\begin{equation}
    \textbf{\emph{u}}^{(k)} = {\rm Enc}^{(k)}(\textbf{\emph{v}}^{(k)})
\end{equation}
We introduce a hidden variable $\textbf{\emph{h}}^{(k)}\in \mathbb{R}^{C\times P\times H \times W}$ and let $\textbf{\emph{h}}^{(0)}$ = $\textbf{\emph{u}}^{(0)}$. After stage 1 ($k\geq 1$), a two-way information exchange occurs between the hidden variable $\textbf{\emph{h}}^{(k-1)}$ and the latent variable $\textbf{\emph{u}}^{(k)}$ at each subsequent stage. This exchange allows $\textbf{\emph{u}}^{(k)}$ to incorporate information from $\textbf{\emph{h}}^{(k-1)}$ that reflects the accumulated knowledge from all previous stages. In turn, $\textbf{\emph{h}}^{(k-1)}$ is updated to $\textbf{\emph{h}}^{(k)}$ based on the information contained within the current stage's latent variable $\textbf{\emph{u}}^{(k)}$. This process is achieved by employing a two-way cross-attention mechanism. Specifically, we leverage the computationally efficient Multi-Dconv Head Transposed Attention (MDTA) from Restormer \cite{9878962}, originally designed for high-resolution images, by adapting it for 3D data processing. To achieve cross-attention, we utilize distinct data sources for the query and key-value components:
\begin{equation}
    \textbf{\emph{q}}_{u}^{(k)} = W_{u}^{q}\textbf{\emph{u}}^{(k)}, 
    \textbf{\emph{k}}_{u}^{(k)} = W_{u}^{k}\textbf{\emph{u}}^{(k)}, 
    \textbf{\emph{v}}_{u}^{(k)} = W_{u}^{v}\textbf{\emph{u}}^{(k)}
\end{equation}
\begin{equation}
    \textbf{\emph{q}}_{h}^{(k)} = W_{h}^{q}\textbf{\emph{h}}^{(k-1)}, 
    \textbf{\emph{k}}_{h}^{(k)} = W_{h}^{k}\textbf{\emph{h}}^{(k-1)}, 
    \textbf{\emph{v}}_{h}^{(k)} = W_{h}^{v}\textbf{\emph{h}}^{(k-1)}
\end{equation}
\begin{equation}
    \begin{aligned}
    \textbf{\emph{u}}'^{(k)} = \textbf{\emph{u}}^{(k)} + {\rm Attn}(\textbf{\emph{q}}_{u}^{(k)}, \textbf{\emph{k}}_{h}^{(k)}, \textbf{\emph{v}}_{h}^{(k)}),\\
    {\rm Attn}(\textbf{\emph{q}}_{u}^{(k)}, \textbf{\emph{k}}_{h}^{(k)}, \textbf{\emph{v}}_{h}^{(k)}) = \textbf{\emph{v}}_{h}^{(k)} \cdot {\rm Softmax}(\textbf{\emph{k}}_{h}^{(k)} \cdot \textbf{\emph{q}}_{u}^{(k)})
    \end{aligned}
\end{equation}
\begin{equation}
    \begin{aligned}
    \textbf{\emph{h}}^{(k)} = \textbf{\emph{h}}^{(k-1)} + {\rm Attn}(\textbf{\emph{q}}_{h}^{(k)}, \textbf{\emph{k}}_{u}^{(k)}, \textbf{\emph{v}}_{u}^{(k)}),\\
    {\rm Attn}(\textbf{\emph{q}}_{h}^{(k)}, \textbf{\emph{k}}_{u}^{(k)}, \textbf{\emph{v}}_{u}^{(k)}) = \textbf{\emph{v}}_{u}^{(k)} \cdot {\rm Softmax}(\textbf{\emph{k}}_{u}^{(k)} \cdot \textbf{\emph{q}}_{h}^{(k)})
    \end{aligned}
\end{equation}
where $W_{(\cdot)}^{q}$, $W_{(\cdot)}^{k}$, and $W_{(\cdot)}^{v}$ denote 3D convolutional layers that project the input feature map into query, key, and value spaces, respectively; $\textbf{\emph{q}}_{(\cdot)}^{(k)}\in \mathbb{R}^{PHW\times C}$, $\textbf{\emph{k}}_{(\cdot)}^{(k)}\in \mathbb{R}^{C\times PHW}$ and $\textbf{\emph{v}}_{(\cdot)}^{(k)}\in \mathbb{R}^{PHW\times C}$ are obtained after reshaping tensors from the original
size $\mathbb{R}^{C\times P\times H \times W}$. After the TWCA module, the $k$-th stage output is generated by the 3D U-net decoder:
\begin{equation}
    \textbf{\emph{x}}^{(k)} = {\rm Dec}^{(k)}(\textbf{\emph{u}}'^{(k)})
\end{equation}

Relying solely on 3D convolutions for image decoding is insufficient, as it leads to block artifacts. A straightforward solution to this issue could be incorporating a 2D U-net at the end for results refinement. In our implementation, we employ a lightweight 2D U-net at the end of BMNet to learn the residual between the output of BMNet's final stage and the ground truth.

\section{Experiment}
\subsection{Datesets and Metrics}
\subsubsection{DOTA-v1.0}
\label{dataset}
We chose the DOTA-v1.0 \cite{Xia_2018_CVPR} dataset to train the BMNet of our compression framework for remote sensing images. The DOTA-v1.0 dataset is a benchmarking large-scale dataset specifically designed for object detection in aerial images. It comprises 2,806 high-resolution images encompassing 15 object categories, including airplanes, ships, storage tanks, sports fields, and bridges. We randomly cropped the 1,411 training set images into non-overlapping patches of size 512$\times$512. From the resulting 41,672 patches, a random sample of 39,588 images was selected for the training set, and the rest 2,084 images were chosen for the testing set. Peak Signal-to-Noise Ratio (PSNR) and Structural Similarity Index Measure (SSIM) were chosen as the evaluation metrics. For consistency with SPI \cite{Song_2023_CVPR, 10.1145/3581783.3612242}, compression is applied solely to the luminance channel (Y) within the YCbCr space. Beyond evaluating reconstruction quality, annotations of DOTA-v1.0 can also be leveraged to assess the mean Average Precision (mAP) of the downstream task object detection under varying compression ratios.
\subsubsection{ISPRS Vaihingen}
\label{ISPRS}
To investigate the impact of compression on a more fine-grained downstream task, we selected the ISPRS Vaihingen \cite{isprs_vaihingen} dataset to test the BMNet trained on DOTA-v1.0 without fine-tuning. The ISPRS Vaihingen dataset is a benchmark for evaluating semantic segmentation algorithms in remote sensing imagery. It comprises 33 high-resolution aerial orthophotos and corresponding digital surface models covering an urban area in Vaihingen, Germany. Only the TOP image tiles were used without the DSM and NDSM. We cropped the images into 31 non-overlapping patches of size 512$\times$512 using ID 34 and 37 as the testing set. The remaining IDs, which were cropped into 3120 non-overlapping patches of size 512$\times$512, were used to train the semantic segmentation model DC-SWin \cite{9681903}. Mean Intersection over Union (mIoU) was chosen as the evaluation metric.
\subsubsection{LandSat8}
\label{landsat8}
We employed the LandSat-8 \cite{landsat8} dataset as an additional test set to assess the generalization capability of the BMI framework for invisible light bands. The LandSat-8 dataset covers a variety of land cover types. It provides global landmass at a spatial resolution of 30 meters (visible, NIR, SWIR); 100 meters (thermal); and 15 meters (panchromatic). Four representative ground type scenes covering vegetation, bare ground, urban, and marine environments were selected and then cropped into 530 non-overlapping patches of size 512$\times$512. SWIR (i.e., band 6) is selected as the test band. PSNR was chosen as the evaluation metric.
\subsubsection{CBSD68}
\label{cbsd}
To facilitate comparison with more SPI approaches, CBSD68 \cite{CBSD68} was selected as an additional test set. It consists of 68 natural color images. Experiments are performed on the luminance channel of YCbCr space. PSNR was chosen as the evaluation metric.

\subsection{Implementation Details}
We configured BMNet with 10 stages and employed a sequence of \{32, 64, 128, 64, 32\} feature dimensions for each 3D U-net within the architecture. 
The loss function is the Mean Squared Error (MSE). We employ the Adam \cite{kingma2014adam} optimizer with a cosine annealing learning rate schedule that incorporates a linear warm-up phase. The initial learning rate is set to $1\times 10^{-6}$, the base learning rate is set to $5\times 10^{-6}$ and the minimum learning rate is set to $5\times 10^{-8}$. Training proceeds for a total of 100 epochs, with the first 5 epochs designated as a warm-up period. Finally, the model is trained on 4 NVIDIA A100 GPUs using a batch size of 4 per GPU. 
Disregarding noise, the optical computing is simulated numerically by performing element-wise multiplication between the image and the binary mask.

\begin{table}[t]
\caption{Comparison of average PSNR (dB) on CBSD68 and DOTA-v1.0 Datasets under varying compression ratios}
\renewcommand{\arraystretch}{1.5}
\setlength{\tabcolsep}{8pt}
\centering
\begin{tabular}{@{}c|c|c|ccc@{}}
\Xhline{1.5pt}
\multirow{2}{*}{\textbf{Dataset}} & \multirow{2}{*}{\textbf{Network}} & \multirow{2}{*}{\textbf{Method}} & \multicolumn{3}{c}{\textbf{CS Ratio}} \\ \Xcline{4-6}{0.4pt}
                         &            &          & 4     &10   & 25          \\ \Xhline{1pt}
\multirow{11}{*}{CBSD68} 
                         & CSNet      & SPI         & 31.12 &27.91   & 25.43    \\
                         & SCSNet     & SPI    & 31.15 &28.02   & 25.37    \\
                         & OPINE-Net  & SPI    & 31.51 &27.82   & 25.20     \\
                         & AMP-Net    & SPI    & 31.80 &27.88   & 25.27    \\
                         & TransCS    & SPI    & 31.74 &27.86   & 25.28     \\
                         & MADUN      & SPI    & 32.27 &28.18   & 25.36     \\
                         & DGUNet     & SPI    & 31.97 &28.13   & 25.45     \\
                         & OCTUF      & SPI    & 32.24 &28.28   & 25.65    \\
                         & SAUNet     & SPI    & 33.67 &29.25   & 26.23   \\ 
                         & \textbf{BMNet(ours)}   & \textbf{BMI}      & \textbf{34.06} &\textbf{29.61}   & \textbf{26.31}         \\ \Xhline{1pt}
\multirow{2}{*}{DOTA-V1.0} 
                        & SAUNet   & SPI     & 44.19 & 39.42   & 35.51   \\
                        & \textbf{BMNet(ours)}   & \textbf{BMI}     & \textbf{43.23} &\textbf{41.81}   & \textbf{35.57}    \\ \Xhline{1.5pt}
\end{tabular}
\label{SPI}
\end{table}

\begin{figure}
    \centering
    \includegraphics[width=0.485\textwidth]{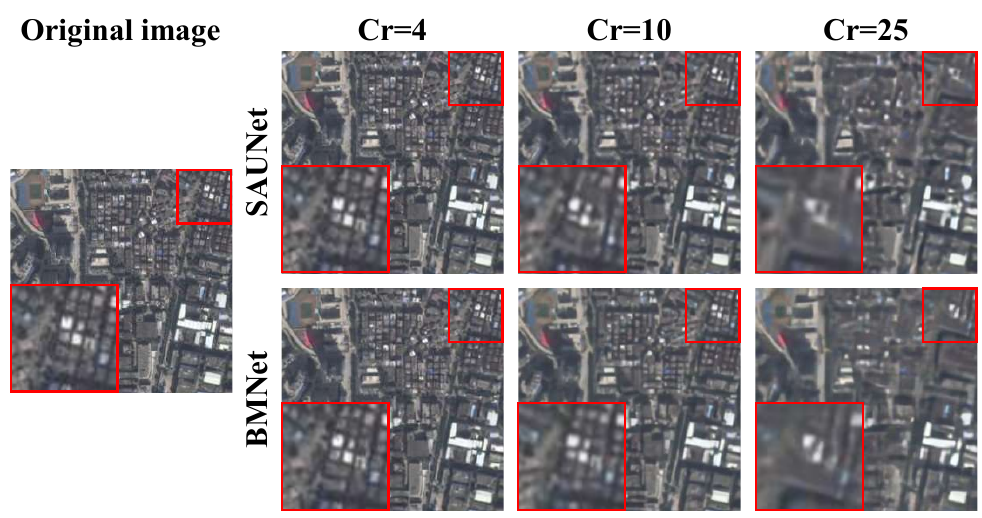}
    \caption{Visualization of the reconstructed images on DOTA-v1.0 at different compression ratios (Cr).}
    \label{bmispi}
\end{figure}

\begin{figure}
    \centering
    \includegraphics[width=0.49\textwidth]{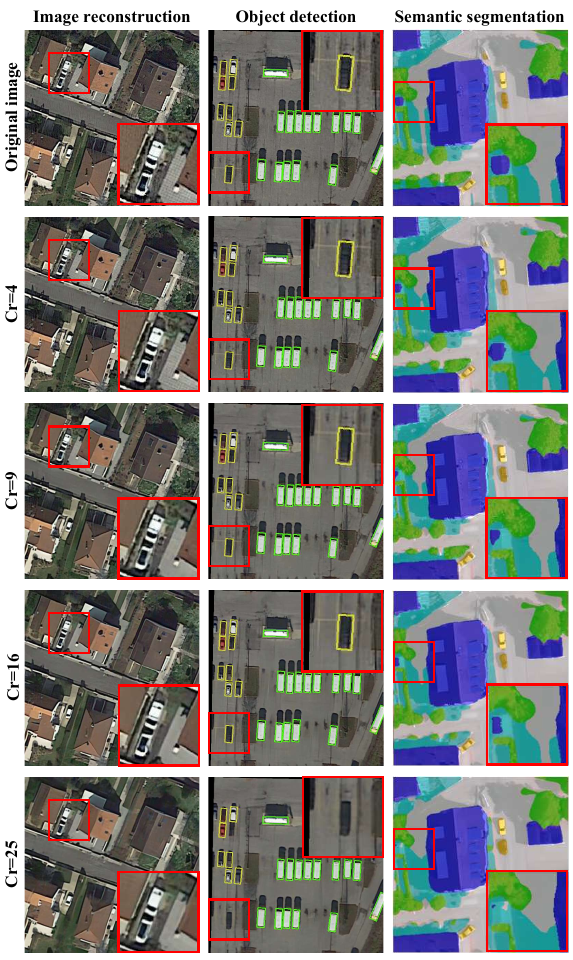}
    \caption{Visualization of different tasks under varying compression ratios (Cr).}
    \label{downvis}
\end{figure}

\subsection{Reconstruction Results Compared with SPI}
In Section \ref{encoding} we analyzed the advantages of BMI over SPI in encoding, in this section we compare the performance of BMI with SPI in terms of decoding. The data utilized for this section are detailed in Sections \ref{dataset} and \ref{cbsd}, respectively.

We first evaluated our BMNet on CBSD68, a benchmark dataset for SPI methods. For a fair comparison, our model also adopted a jointly-trained measurement matrix here. We subsequently evaluated the performance of BMNet and the SOTA SPI decoding network, SAUNet \cite{10.1145/3581783.3612242}, on DOTA-v1.0 dataset. Due to the high computational complexity of mainstream SPI decoding algorithms, particularly when applied to high-resolution images, replicating them on DOTA-v1.0 dataset would be prohibitively expensive (see Section \ref{highre}). For experiments on DOTA-v1.0 in this section, we resize the images to a resolution of 256$\times$256. We began by training the models at a compression ratio of 10. The pre-trained models were then utilized as foundations to fine-tune models for compression ratios of 4 and 25.

The visual results are displayed in Fig. \ref{bmispi}, and the corresponding average PSNR values are summarized in Table \ref{SPI}. BMNet exhibits performance comparable to or exceeding that of SAUNet, the SOTA SPI method. This result is particularly encouraging as it demonstrates the potential for BMI to replace SPI as a solution for on-orbit remote sensing image compression with both lower encoding complexity and superior decoding performance.

\begin{figure*}[t]
    \centering
    \includegraphics[width=0.95\textwidth]{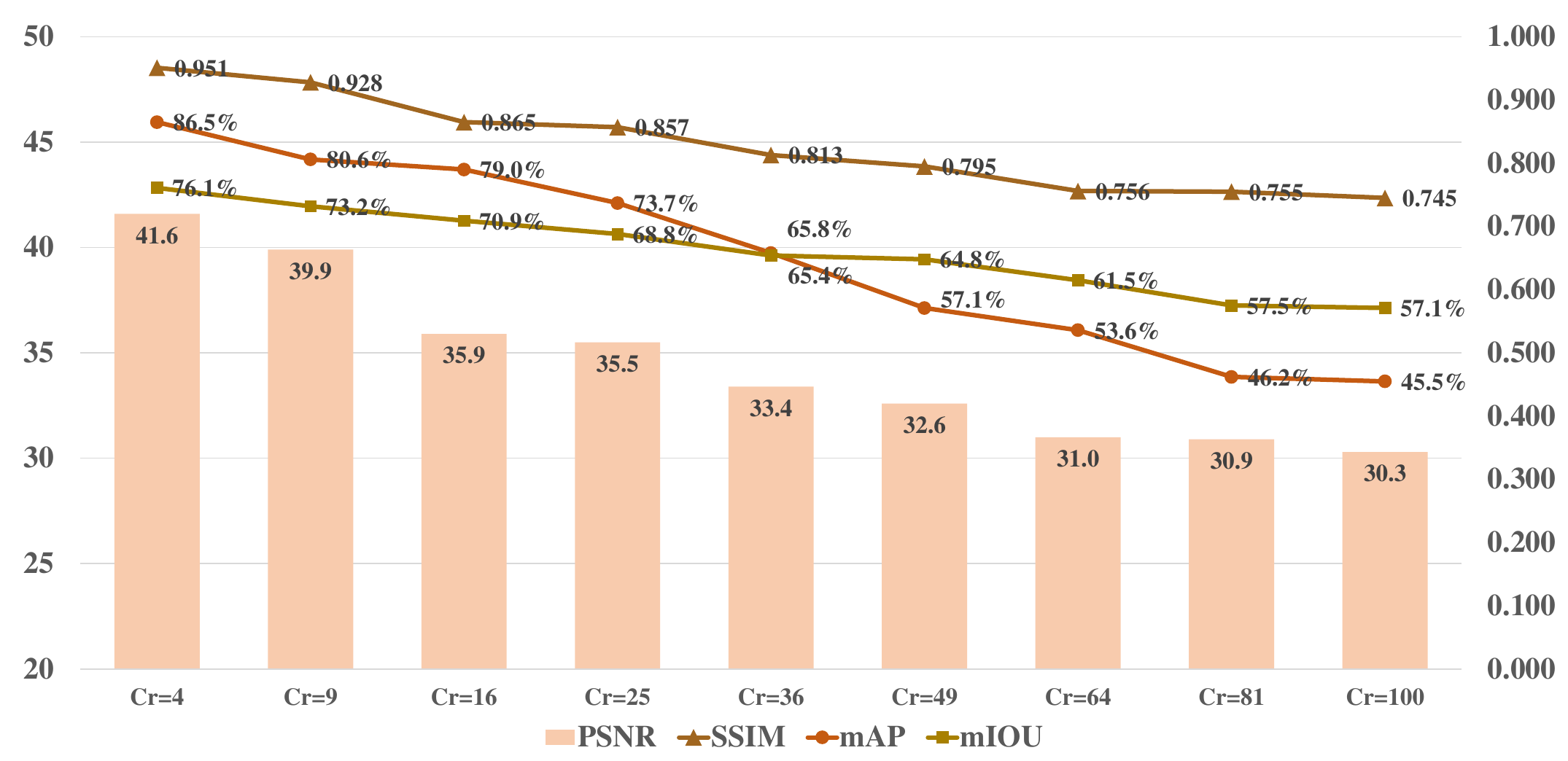}
    \caption{Comprehensive performance of BMNet under varying compression ratios (Cr): PSNR, SSIM for image reconstruction (DOTA-v1.0), mAP for object detection (DOTA-v1.0) and mIoU for semantic segmentation (Vaihingen).}
    \label{downstream}
\end{figure*}

\subsection{Reconstruction Results on Downstream Tasks}
For lossy compression methods, it is essential to analyze the impact of compression ratio on the efficacy of subsequent downstream tasks. Most compressed sensing methods have primarily concentrated on minimizing reconstruction error. However the quality of reconstruction, as measured by metrics like mean squared error (MSE), might not always align directly with the performance requirements of a specific application. In this section, a baseline will be established for the tasks of object detection and semantic segmentation, serving as a benchmark for evaluating the effectiveness of future remote sensing image compression methods based on compressed sensing. The data utilized for these experiments are detailed in Sections \ref{dataset} and \ref{ISPRS}, respectively.

To obtain decoding networks for different compression ratios, we first establish a base BMNet trained with a compression ratio of 16 on DOTA-v1.0. Subsequently, we leverage this pre-trained model for fine-tuning additional models with compression ratios of 4, 9, 25, 36, 49, 64, 81 and 100. For the object detection task, we first evaluated the pre-trained YOLO-v5s \cite{yolo} model on the uncompressed testing set, achieving a mAP of 86.5\%. Subsequently, we applied our method to compress and then reconstruct the testing set under varying compression ratios. The reconstructed testing set was then fed into the YOLO model for detection. For the semantic segmentation task, We first trained the DC-Swin \cite{9681903} model and achieved a mIoU of 78.6\% on the uncompressed testing set. Then the testing set, which has undergone compression and subsequent reconstruction via the BMI framework, is once again employed to evaluate the performance of the trained DC-Swin model at varying compression ratios.

Fig. \ref{downvis} visualizes the reconstruction restults. The average PSNR, SSIM, mAP for DOTA-v1.0 dataset and the average mIoU for Vaihingen dataest are presented in Fig. \ref{downstream}. Negligible semantic information is lost at a compression ratio of 4. The reduction in the performance of downstream tasks is constrained to a maximum of 7.5\% for compression ratios under 16. These results provide insights into the limitations of the compression ratio. We acknowledge that more powerful YOLO and DC-Swin models or task-driven training could improve detection and segmentation performance on the reconstructed images, however these approaches are beyond this paper's scope.

\begin{table}[]
\caption{Average PSNR (dB) of BMNet on LandSat8}
\centering
\renewcommand{\arraystretch}{1.5}
\begin{tabular}{@{}c|cccc@{}}
\Xhline{1.5pt}
\multirow{2}{*}{\textbf{Dataset}} & \multicolumn{4}{c}{\textbf{CS Ratio}} \\ \Xcline{2-5}{0.4pt}
                         & 4     & 9     & 16    & 25   \\ \Xhline{1pt}
LandSat8                 & 47.0  & 45.5  & 41.9  & 41.8 \\ \Xhline{1.5pt}
\end{tabular}
\label{swirpsnr}
\end{table}

\begin{figure}
    \centering
    \includegraphics[width=0.485\textwidth]{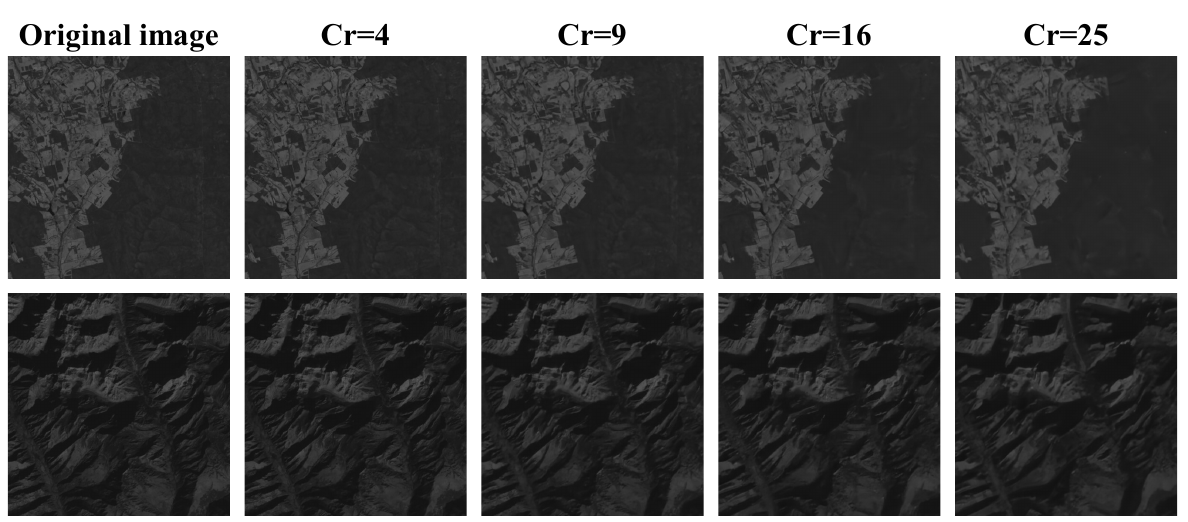}
    \caption{Visualization of the reconstructed SWIR band at different compression ratios (Cr).}
    \label{SWIR}
\end{figure}

\subsection{Reconstruction Results on Invisible Band}
For remote sensing scenarios, it is often necessary to deal with invisible spectral bands rather than just grey scale or RGB images. This section investigates the generalizability of BMNet to SWIR spectral band. The data utilized here are detailed in Section \ref{landsat8}.

Fig. \ref{SWIR} presents the visual effects and Table \ref{swirpsnr} illustrates average PSNR values of the reconstruction results. BMNet demonstrates strong generalization capabilities and effectively reconstructed the SWIR band.

\subsection{Prototype Designing and Testing}
\begin{figure}
    \centering
    \subfloat[]{\includegraphics[width=0.36\textwidth]{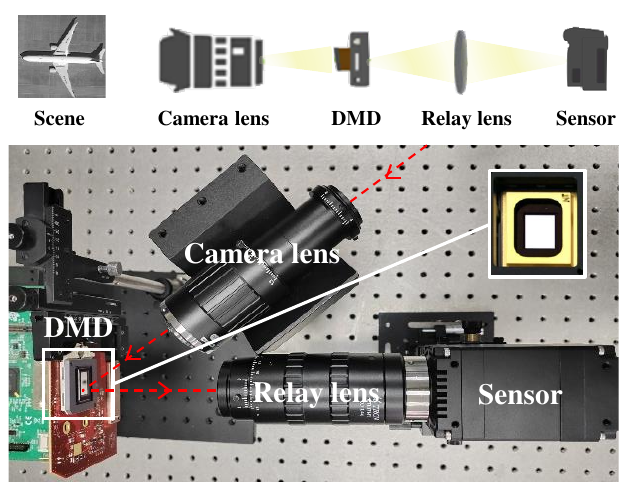}}\\  
    \subfloat[]{\includegraphics[width=0.48\textwidth]{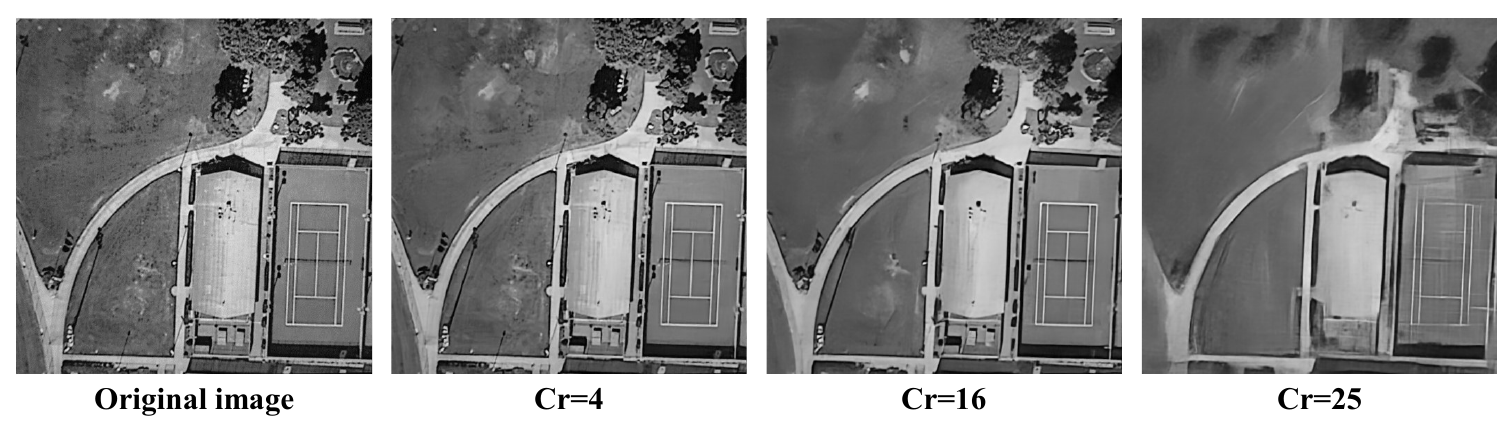}}
    \caption{Real-world testing of the prototype. (a) The digital acquisition unit. (b) Reconstruction result.}
    \label{proto}
\end{figure}

We implemented the simulation process in a physical system by designing a digital acquisition unit hardware system based on the Metacam from \cite{RN53}. Given its flexibility, the DMD \cite{Hu:21} is retained in the prototype to accommodate experimentation with different measurement matrix configurations. Note that in practical applications, the DMD can be substituted with a photomask for high-resolution imaging.
Fig. \ref{proto} (a) depicts the physical layout of the entire optical imaging system. The digital acquisition unit comprises an imaging lens assembly, an optical coding unit, and a relay lens assembly. The scene is imaged onto a virtual plane through the camera lens (CHIOPT HC3505A) for original data acquisition. The relay lens (Thorlabs MAP10100100-A) focuses the target scene image onto the DMD (ViALUX V-9001, 2560×1600 resolution, 7.6 $\mu m$ pixel pitch), forming a primary image. The DMD modulates the data by adjusting the amplitude of the light, achieving instantaneous encoding. The reflected light from the DMD is focused onto the image sensor (FLIR GS3-U3-120S6M-C, 4242×2830 resolution, 3.1 $\mu m$ pixel pitch) by the zoom lens (Utron VTL0714V). The relay lens group is used to transmit the encoded data to the photosensitive surface of the sensor, so that one DMD mirror matches one image sensor pixel, ensuring precise alignment of the DMD and the image sensor.
Notably, this system exhibits a flexible, plug-and-play characteristic. The degree of hardware integration within the application has the potential for significant enhancement. Furthermore, the system possesses the adaptability to accommodate various compression ratios, as the compression ratio depends solely on electrical computations.

\begin{figure}
    \centering 
    \subfloat[]{\includegraphics[width=0.225\textwidth]{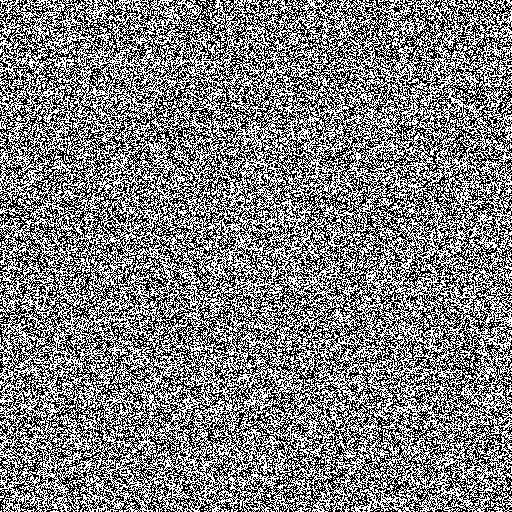}}\quad
    \subfloat[]{\includegraphics[width=0.225\textwidth]{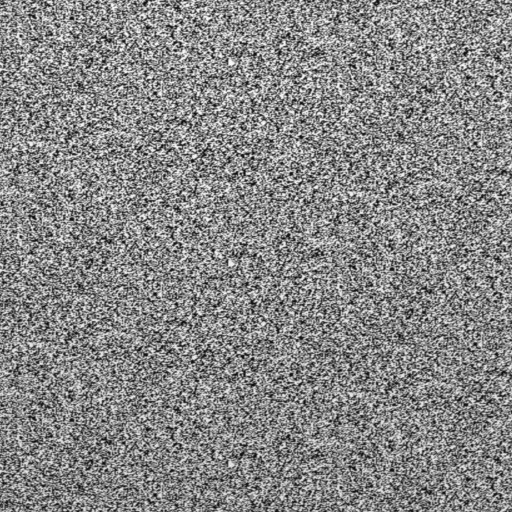}}
    \caption{Comparison of masks in different environments. (a) Simulated environment mask. (b) Real environment mask.}
    \label{realmask}
\end{figure}

To validate the reconstruction algorithm's performance in real-world scenarios, we built a sandbox to simulate remote sensing scenarios within a natural light environment. The projected data was captured by the digital acquisition unit to form measurements, and the encoded measurements was then fed into BMNet for reconstruction.
Due to the diagonal hinge design of the DMD's digital micromirrors, their rotation for modulation occurs along an axis tilted 45° relative to the array orientation \cite{10.1063/5.0040424}. This results in a 45° image rotation after processing by the digital acquisition unit. To address this issue, we employed image zooming to magnify the projected image and subsequent cropping to achieve a captured image size of 512×512, satisfying the required dimensions for reconstruction.
Furthermore, we observed that due to the use of lenses in the alignment process, the calibrated masks will deviate from the binary encoding used for the simulation. To address this problem, we fine-tuned the model using the mask obtained in the real environment. Specifically, a uniform illumination source, provided by an integrating sphere, illuminates the entrance pupil. The binary mask is then loaded onto the DMD. The resulting coded light pattern serves as the actual experimental mask in real-world testing, as shown in Fig. \ref{realmask}.

The reconstruction results for a compression ratio of 4, 16 and 25 are shown in Fig. \ref{proto} (b). The image was successfully encoded by the digital acquisition unit and then reconstructed by BMNet. The reconstruction results exhibit discrepancies compared to the simulation. These deviations can be attributed to various factors including fluctuations in optical field intensity, imperfections in relay lens calibration, inherent system errors, and unavoidable error in artificial adjustment of the image resolution, as was observed in \cite{10376842}. Considering sensor response could potentially mitigate this issue, however it falls outside the scope of this paper.

\subsection{Ablation Study}
\begin{table}[]
\renewcommand{\arraystretch}{1.5}
\caption{Ablation study of the proposed gated 3D convolution and TWCA at a compression ratio of 16}
\centering
\begin{tabular}{@{}ccc|c@{}}
\Xhline{1.5pt}
Gated Conv. & TWCA & Params. (M) & PSNR (dB) \\ \Xhline{1pt}
\scalebox{0.75}{\usym{2613}}           & \scalebox{0.75}{\usym{2613}}    & 22.23   & 34.9 \\
\checkmark           & \scalebox{0.75}{\usym{2613}}    & 22.51   & 35.1 \\
\scalebox{0.75}{\usym{2613}}           & \checkmark    & 33.17   & 35.8  \\
\checkmark           & \checkmark    &  33.45    & 35.9 \\ \Xhline{1.5pt}
\end{tabular}
\label{ablation}
\end{table}

\begin{table}
\renewcommand{\arraystretch}{1.5}
\caption{Comparison of decoding speed (ms/image) between SPI and BMI at a compression ratio of 16}
\centering
\begin{tabular}{@{}c|ccc@{}}
\Xhline{1.5pt}
\multirow{2}{*}{\textbf{Method}} & \multicolumn{3}{c}{\textbf{Resolution}} \\ \Xcline{2-4}{0.4pt}
                        & 256x256  & 512x512 & 1024x1024 \\ \Xhline{1pt}
SAUNet                     & 103.1     & 284.9    & 1326.6      \\
BMNet                     & 54.9     & 79.9    & 376.8      \\ \Xhline{1.5pt}
\end{tabular}
\label{speed}
\end{table}

To validate the effectiveness of the proposed gated 3D convolution and TWCA, we conducted ablation experiments on each module under a compression ratio of 16, as presented in Table \ref{ablation}. The gating mechanism gains 0.2dB with minimal parameter increase. Incorporating cross-attention mechanisms for information interaction at different stages yielded a 0.9dB performance boost. In combination, these two techniques deliver a 1dB improvement.
\subsection{Computational Analysis}
\subsubsection{High-resolution Decoding}
\label{highre}
A primary resolution of 512$\times$512 was employed in this paper to streamline the evaluation of various measurement matrices utilizing DMDs and accelerate training. This section presents a comparative analysis of decoding speeds for BMI and SPI at increasing resolutions, using a compression ratio of 16 as shown in Table \ref{speed}. To evaluate model throughput, we measured the average inference time required by the network to process a grayscale image of varying resolutions on a 4090 GPU. The measurement was repeated 100 times to account for potential variations. BMNet exhibits an inference speed that is 2$ \sim $4 times faster than SAUNet, the SOTA SPI method. This further illustrates the advantages of BMI, It is not only more suited for encoding high-resolution images, but it also excels in decoding processes.

\subsubsection{Advantages over JPEG}
\begin{figure}
    \centering
    \includegraphics[width=0.485\textwidth]{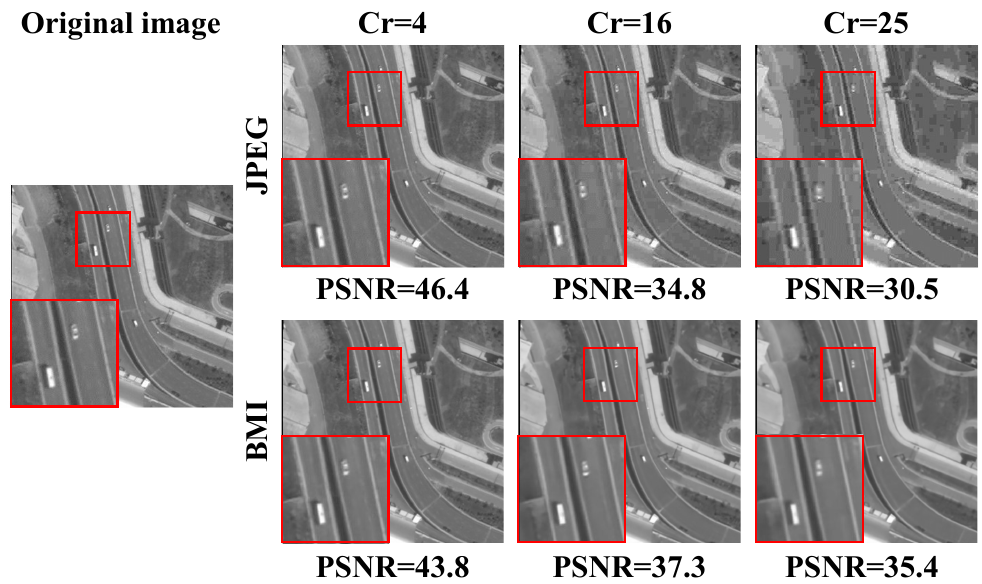}
    \caption{Decoding performance of JPEG and BMI at 512x512 resolution under varying compression ratios (Cr).}
    \label{jpeg_de}
\end{figure}

JPEG has long been the industry standard for image compression. In this section, JPEG, as a representative of traditional compression methods, is compared to BMI to highlight its advantages in both encoding and decoding efficiency. Table \ref{jpeg-en} compares the encoding speeds of JPEG and BMI, where JPEG is implemented using the OpenCV library\cite{opencv}. As resolution increases, BMI exhibits significantly slower growth in encoding time compared to JPEG, making it particularly advantageous for large-scale remote sensing data. Fig. \ref{jpeg_de} demonstrates the decoding performance of JPEG and BMI across various compression ratios. Notably, BMI exhibits superior performance at high compression ratios, further illustrating its advantages.

\begin{table}
\renewcommand{\arraystretch}{1.5}
\caption{Comparison of encoding speed (ms/image) between JPEG and BMI at a compression ratio of 16}
\centering
\begin{tabular}{@{}c|ccc@{}}
\Xhline{1.5pt}
\multirow{2}{*}{\textbf{Method}} & \multicolumn{3}{c}{\textbf{Resolution}} \\ \Xcline{2-4}{0.4pt}
                        & 512x512  & 2048x2048 & 8192x8192 \\ \Xhline{1pt}
JPEG                     & 3.11     & 54.88    & 815.31      \\
BMI                     & 0.03     & 0.64    & 14.46      \\ \Xhline{1.5pt}
\end{tabular}
\label{jpeg-en}
\end{table}

\section{Conclusion}
Leveraging the principles of compressed sensing, we propose a novel ultra-low encoding complexity paradigm for remote sensing image compression, termed Block Modulated Imaging (BMI). By exploiting the capabilities of optical modulation, this approach offers extremely low encoding complexity, making it highly attractive for resource-constrained platforms with limited computational and storage capabilities. We further propose a deep unfolding network leveraging gated 3D convolutions and incorporate a cross-attention mechanism to facilitate information exchange across different stages. BMI exhibits superior performance to SPI in both encoding and decoding processes. Experiments verify the effectiveness of our method, the paracticality and functionality of which is also verified by designing and testing a hardware prototype. 
We hope that our proposed formula and corresponding hardware can open a new venue and motivate more research in the field of high-resolution scene capturing and processing.

\section*{Acknowledgments}
This work was supported in part by the National Natural Science Foundation of China under grant U23B2011, 62102069, U20B2063 and 62220106008.




\bibliographystyle{IEEEtran}
\bibliography{IEEEabrv, mylib}

\begin{thebibliography}{10}
\providecommand{\url}[1]{#1}
\csname url@samestyle\endcsname
\providecommand{\newblock}{\relax}
\providecommand{\bibinfo}[2]{#2}
\providecommand{\BIBentrySTDinterwordspacing}{\spaceskip=0pt\relax}
\providecommand{\BIBentryALTinterwordstretchfactor}{4}
\providecommand{\BIBentryALTinterwordspacing}{\spaceskip=\fontdimen2\font plus
\BIBentryALTinterwordstretchfactor\fontdimen3\font minus \fontdimen4\font\relax}
\providecommand{\BIBforeignlanguage}[2]{{%
\expandafter\ifx\csname l@#1\endcsname\relax
\typeout{** WARNING: IEEEtran.bst: No hyphenation pattern has been}%
\typeout{** loaded for the language `#1'. Using the pattern for}%
\typeout{** the default language instead.}%
\else
\language=\csname l@#1\endcsname
\fi
#2}}
\providecommand{\BIBdecl}{\relax}
\BIBdecl

\bibitem{10239453}
S.~Dang, Z.~Xia, X.~Jiang, S.~Gui, and X.~Feng, ``Inclusive consistency-based quantitative decision-making framework for incremental automatic target recognition,'' \emph{IEEE Transactions on Geoscience and Remote Sensing}, vol.~61, pp. 1--14, 2023.

\bibitem{8350339}
D.~S. Plotnick and T.~M. Marston, ``Utilization of aspect angle information in synthetic aperture images,'' \emph{IEEE Transactions on Geoscience and Remote Sensing}, vol.~56, no.~9, pp. 5424--5432, 2018.

\bibitem{4215094}
S.~Voigt, T.~Kemper, T.~Riedlinger, R.~Kiefl, K.~Scholte, and H.~Mehl, ``Satellite image analysis for disaster and crisis-management support,'' \emph{IEEE Transactions on Geoscience and Remote Sensing}, vol.~45, no.~6, pp. 1520--1528, 2007.

\bibitem{7565634}
M.~Chi, A.~Plaza, J.~A. Benediktsson, Z.~Sun, J.~Shen, and Y.~Zhu, ``Big data for remote sensing: Challenges and opportunities,'' \emph{Proceedings of the IEEE}, vol. 104, no.~11, pp. 2207--2219, 2016.

\bibitem{10439079}
S.~Shi, H.~Wang, and X.~Ma, ``Key issues on application of remote sensing big data (rsbd): Concepts, scenarios, and challenges,'' in \emph{2023 2nd International Conference on Cloud Computing, Big Data Application and Software Engineering (CBASE)}, 2023, pp. 238--245.

\bibitem{9144368}
A.~Preethy~Byju, G.~Sumbul, B.~Demir, and L.~Bruzzone, ``Remote-sensing image scene classification with deep neural networks in jpeg 2000 compressed domain,'' \emph{IEEE Transactions on Geoscience and Remote Sensing}, vol.~59, no.~4, pp. 3458--3472, 2021.

\bibitem{RN52}
\BIBentryALTinterwordspacing
S.~Elakkiya and K.~S. Thivya, ``Comprehensive review on lossy and lossless compression techniques,'' \emph{Journal of The Institution of Engineers (India): Series B}, vol. 103, no.~3, pp. 1003--1012, 2022. [Online]. Available: \url{https://doi.org/10.1007/s40031-021-00686-3}
\BIBentrySTDinterwordspacing

\bibitem{HUSSAIN201844}
\BIBentryALTinterwordspacing
A.~Hussain, A.~Al-Fayadh, and N.~Radi, ``Image compression techniques: A survey in lossless and lossy algorithms,'' \emph{Neurocomputing}, vol. 300, pp. 44--69, 2018. [Online]. Available: \url{https://www.sciencedirect.com/science/article/pii/S0925231218302935}
\BIBentrySTDinterwordspacing

\bibitem{6626242}
X.~Chen and X.~Xu, ``A highly efficient rate control algorithm for jpeg2000 images,'' \emph{IEEE Transactions on Consumer Electronics}, vol.~59, no.~3, pp. 587--591, 2013.

\bibitem{7026133}
F.~Aulí-Llinàs, ``Highly efficient, low complexity arithmetic coder for jpeg2000,'' in \emph{2014 IEEE International Conference on Image Processing (ICIP)}, 2014, pp. 5601--5605.

\bibitem{2017.0469}
\BIBentryALTinterwordspacing
T.~Ma, ``Low-complexity and efficient image coder/decoder with quad-tree search model for embedded computing platforms,'' \emph{IET Image Processing}, vol.~12, no.~2, pp. 235--242, 2018. [Online]. Available: \url{https://ietresearch.onlinelibrary.wiley.com/doi/abs/10.1049/iet-ipr.2017.0469}
\BIBentrySTDinterwordspacing

\bibitem{malathkar2020near}
N.~V. Malathkar and S.~K. Soni, ``A near lossless and low complexity image compression algorithm based on fixed threshold dpcm for capsule endoscopy,'' \emph{Multimedia Tools and Applications}, vol.~79, no.~11, pp. 8145--8160, 2020.

\bibitem{1614066}
D.~Donoho, ``Compressed sensing,'' \emph{IEEE Transactions on Information Theory}, vol.~52, no.~4, pp. 1289--1306, 2006.

\bibitem{1697831}
C.~Shannon, ``Communication in the presence of noise,'' \emph{Proceedings of the IRE}, vol.~37, no.~1, pp. 10--21, 1949.

\bibitem{4472247}
M.~F. Duarte, M.~A. Davenport, D.~Takhar, J.~N. Laska, T.~Sun, K.~F. Kelly, and R.~G. Baraniuk, ``Single-pixel imaging via compressive sampling,'' \emph{IEEE Signal Processing Magazine}, vol.~25, no.~2, pp. 83--91, 2008.

\bibitem{4752747}
J.~Ma, ``Single-pixel remote sensing,'' \emph{IEEE Geoscience and Remote Sensing Letters}, vol.~6, no.~2, pp. 199--203, 2009.

\bibitem{52362346234}
------, ``A single-pixel imaging system for remote sensing by two-step iterative curvelet thresholding,'' \emph{Geoscience and Remote Sensing Letters, IEEE}, vol.~6, pp. 676 -- 680, 11 2009.

\bibitem{adfdsadfb}
\BIBentryALTinterwordspacing
Z.~Wang and J.~Zhu, ``Single-pixel compressive imaging based on motion compensation,'' \emph{IET Image Processing}, vol.~12, no.~12, pp. 2283--2291, 2018. [Online]. Available: \url{https://ietresearch.onlinelibrary.wiley.com/doi/abs/10.1049/iet-ipr.2018.5741}
\BIBentrySTDinterwordspacing

\bibitem{8588994}
J.~Li, Y.~Fu, G.~Li, and Z.~Liu, ``Remote sensing image compression in visible/near-infrared range using heterogeneous compressive sensing,'' \emph{IEEE Journal of Selected Topics in Applied Earth Observations and Remote Sensing}, vol.~11, no.~12, pp. 4932--4938, 2018.

\bibitem{10.1145/103085.103089}
\BIBentryALTinterwordspacing
G.~K. Wallace, ``The jpeg still picture compression standard,'' \emph{Commun. ACM}, vol.~34, no.~4, p. 30–44, apr 1991. [Online]. Available: \url{https://doi.org/10.1145/103085.103089}
\BIBentrySTDinterwordspacing

\bibitem{952804}
A.~Skodras, C.~Christopoulos, and T.~Ebrahimi, ``The jpeg 2000 still image compression standard,'' \emph{IEEE Signal Processing Magazine}, vol.~18, no.~5, pp. 36--58, 2001.

\bibitem{elkdhssdf}
\BIBentryALTinterwordspacing
Z.~Zhang and J.~Zhong, ``Three-dimensional single-pixel imaging with far fewer measurements than effective image pixels,'' \emph{Opt. Lett.}, vol.~41, no.~11, pp. 2497--2500, Jun 2016. [Online]. Available: \url{https://opg.optica.org/ol/abstract.cfm?URI=ol-41-11-2497}
\BIBentrySTDinterwordspacing

\bibitem{Monin2021}
\BIBentryALTinterwordspacing
S.~Monin, E.~Hahamovich, and A.~Rosenthal, ``Single-pixel imaging of dynamic objects using multi-frame motion estimation,'' \emph{Scientific Reports}, vol.~11, no.~1, p. 7712, Apr 2021. [Online]. Available: \url{https://doi.org/10.1038/s41598-021-83810-z}
\BIBentrySTDinterwordspacing

\bibitem{Li:21}
\BIBentryALTinterwordspacing
X.~Li, Y.~Yin, W.~He, X.~Liu, Q.~Tang, and X.~Peng, ``New strategy for high-dimensional single-pixel imaging,'' \emph{Opt. Express}, vol.~29, no.~22, pp. 36\,675--36\,688, Oct 2021. [Online]. Available: \url{https://opg.optica.org/oe/abstract.cfm?URI=oe-29-22-36675}
\BIBentrySTDinterwordspacing

\bibitem{Stojek:22}
\BIBentryALTinterwordspacing
R.~Stojek, A.~Pastuszczak, P.~Wr\'{o}bel, and R.~Koty\'{n}ski, ``Single pixel imaging at high pixel resolutions,'' \emph{Opt. Express}, vol.~30, no.~13, pp. 22\,730--22\,745, Jun 2022. [Online]. Available: \url{https://opg.optica.org/oe/abstract.cfm?URI=oe-30-13-22730}
\BIBentrySTDinterwordspacing

\bibitem{xue2022block}
Y.~Xue, S.~Zheng, W.~Tahir, Z.~Wang, H.~Zhang, Z.~Meng, L.~Tian, and X.~Yuan, ``Block modulating video compression: an ultra low complexity image compression encoder for resource limited platforms,'' \emph{arXiv preprint arXiv:2205.03677}, 2022.

\bibitem{Yuan_2020_CVPR}
X.~Yuan, Y.~Liu, J.~Suo, and Q.~Dai, ``Plug-and-play algorithms for large-scale snapshot compressive imaging,'' in \emph{Proceedings of the IEEE/CVF Conference on Computer Vision and Pattern Recognition (CVPR)}, June 2020.

\bibitem{meng2020gap}
Z.~Meng, S.~Jalali, and X.~Yuan, ``Gap-net for snapshot compressive imaging,'' \emph{arXiv preprint arXiv:2012.08364}, 2020.

\bibitem{7153923}
S.~Zhou, C.~Deng, B.~Zhao, Y.~Xia, Q.~Li, and Z.~Chen, ``Remote sensing image compression: A review,'' in \emph{2015 IEEE International Conference on Multimedia Big Data}, April 2015, pp. 406--410.

\bibitem{7570236}
M.~Conoscenti, R.~Coppola, and E.~Magli, ``Constant snr, rate control, and entropy coding for predictive lossy hyperspectral image compression,'' \emph{IEEE Transactions on Geoscience and Remote Sensing}, vol.~54, no.~12, pp. 7431--7441, 2016.

\bibitem{7935537}
J.~Bartrina-Rapesta, I.~Blanes, F.~Aulí-Llinàs, J.~Serra-Sagristà, V.~Sanchez, and M.~W. Marcellin, ``A lightweight contextual arithmetic coder for on-board remote sensing data compression,'' \emph{IEEE Transactions on Geoscience and Remote Sensing}, vol.~55, no.~8, pp. 4825--4835, 2017.

\bibitem{13-6001-5_42}
S.~K. Gunasheela and H.~S. Prasantha, ``Compressed sensing for image compression: Survey of algorithms,'' in \emph{Emerging Research in Computing, Information, Communication and Applications}, N.~R. Shetty, L.~M. Patnaik, H.~C. Nagaraj, P.~N. Hamsavath, and N.~Nalini, Eds.\hskip 1em plus 0.5em minus 0.4em\relax Singapore: Springer Singapore, 2019, pp. 507--517.

\bibitem{8898608}
J.~Bobak, H.~Alqadah, M.~Nurnberger, S.~Rudolph, and D.~Truesdale, ``Microwave single pixel imager (mspi) overview and imaging algorithm,'' in \emph{IGARSS 2019 - 2019 IEEE International Geoscience and Remote Sensing Symposium}, 2019, pp. 8837--8840.

\bibitem{NIPS2016_1679091c}
\BIBentryALTinterwordspacing
y.~yang, J.~Sun, H.~Li, and Z.~Xu, ``Deep admm-net for compressive sensing mri,'' in \emph{Advances in Neural Information Processing Systems}, D.~Lee, M.~Sugiyama, U.~Luxburg, I.~Guyon, and R.~Garnett, Eds., vol.~29.\hskip 1em plus 0.5em minus 0.4em\relax Curran Associates, Inc., 2016. [Online]. Available: \url{https://proceedings.neurips.cc/paper_files/paper/2016/file/1679091c5a880faf6fb5e6087eb1b2dc-Paper.pdf}
\BIBentrySTDinterwordspacing

\bibitem{Ma_2019_ICCV}
J.~Ma, X.-Y. Liu, Z.~Shou, and X.~Yuan, ``Deep tensor admm-net for snapshot compressive imaging,'' in \emph{Proceedings of the IEEE/CVF International Conference on Computer Vision (ICCV)}, October 2019.

\bibitem{9363502}
X.~Yuan, D.~J. Brady, and A.~K. Katsaggelos, ``Snapshot compressive imaging: Theory, algorithms, and applications,'' \emph{IEEE Signal Processing Magazine}, vol.~38, no.~2, pp. 65--88, 2021.

\bibitem{9711280}
\BIBentryALTinterwordspacing
Z.~Wt, J.~Zhangt, and C.~Mou, ``Dense deep unfolding network with 3d-cnn prior for snapshot compressive imaging,'' in \emph{2021 IEEE/CVF International Conference on Computer Vision (ICCV)}.\hskip 1em plus 0.5em minus 0.4em\relax Los Alamitos, CA, USA: IEEE Computer Society, oct 2021, pp. 4872--4881. [Online]. Available: \url{https://doi.ieeecomputersociety.org/10.1109/ICCV48922.2021.00485}
\BIBentrySTDinterwordspacing

\bibitem{10377358}
S.~Zheng and X.~Yuan, ``Unfolding framework with prior of convolution-transformer mixture and uncertainty estimation for video snapshot compressive imaging,'' in \emph{2023 IEEE/CVF International Conference on Computer Vision (ICCV)}, 2023, pp. 12\,692--12\,703.

\bibitem{bi}
\BIBentryALTinterwordspacing
Z.~Cheng, R.~Lu, Z.~Wang, H.~Zhang, B.~Chen, Z.~Meng, and X.~Yuan, ``Birnat: Bidirectional recurrent neural networks with adversarial training for video snapshot compressive imaging.''\hskip 1em plus 0.5em minus 0.4em\relax Berlin, Heidelberg: Springer-Verlag, 2020, p. 258–275. [Online]. Available: \url{https://doi.org/10.1007/978-3-030-58586-0_16}
\BIBentrySTDinterwordspacing

\bibitem{Cheng_2021_CVPR}
Z.~Cheng, B.~Chen, G.~Liu, H.~Zhang, R.~Lu, Z.~Wang, and X.~Yuan, ``Memory-efficient network for large-scale video compressive sensing,'' in \emph{Proceedings of the IEEE/CVF Conference on Computer Vision and Pattern Recognition (CVPR)}, June 2021, pp. 16\,246--16\,255.

\bibitem{MAL-016}
\BIBentryALTinterwordspacing
S.~Boyd, N.~Parikh, E.~Chu, B.~Peleato, and J.~Eckstein, ``Distributed optimization and statistical learning via the alternating direction method of multipliers,'' \emph{Foundations and Trends® in Machine Learning}, vol.~3, no.~1, pp. 1--122, 2011. [Online]. Available: \url{http://dx.doi.org/10.1561/2200000016}
\BIBentrySTDinterwordspacing

\bibitem{gap}
\BIBentryALTinterwordspacing
X.~Liao, H.~Li, and L.~Carin, ``Generalized alternating projection for weighted-$\ell_{2,1}$ minimization with applications to model-based compressive sensing,'' \emph{SIAM Journal on Imaging Sciences}, vol.~7, no.~2, pp. 797--823, 2014. [Online]. Available: \url{https://doi.org/10.1137/130936658}
\BIBentrySTDinterwordspacing

\bibitem{Yu_2019_ICCV}
J.~Yu, Z.~Lin, J.~Yang, X.~Shen, X.~Lu, and T.~S. Huang, ``Free-form image inpainting with gated convolution,'' in \emph{Proceedings of the IEEE/CVF International Conference on Computer Vision (ICCV)}, October 2019.

\bibitem{Song_2023_CVPR}
J.~Song, C.~Mou, S.~Wang, S.~Ma, and J.~Zhang, ``Optimization-inspired cross-attention transformer for compressive sensing,'' in \emph{Proceedings of the IEEE/CVF Conference on Computer Vision and Pattern Recognition (CVPR)}, June 2023, pp. 6174--6184.

\bibitem{10.1145/3474085.3475562}
\BIBentryALTinterwordspacing
J.~Song, B.~Chen, and J.~Zhang, ``Memory-augmented deep unfolding network for compressive sensing,'' in \emph{Proceedings of the 29th ACM International Conference on Multimedia}, ser. MM '21.\hskip 1em plus 0.5em minus 0.4em\relax New York, NY, USA: Association for Computing Machinery, 2021, p. 4249–4258. [Online]. Available: \url{https://doi.org/10.1145/3474085.3475562}
\BIBentrySTDinterwordspacing

\bibitem{9878962}
S.~W. Zamir, A.~Arora, S.~Khan, M.~Hayat, F.~S. Khan, and M.~Yang, ``Restormer: Efficient transformer for high-resolution image restoration,'' in \emph{2022 IEEE/CVF Conference on Computer Vision and Pattern Recognition (CVPR)}, 2022, pp. 5718--5729.

\bibitem{Xia_2018_CVPR}
G.-S. Xia, X.~Bai, J.~Ding, Z.~Zhu, S.~Belongie, J.~Luo, M.~Datcu, M.~Pelillo, and L.~Zhang, ``Dota: A large-scale dataset for object detection in aerial images,'' in \emph{Proceedings of the IEEE Conference on Computer Vision and Pattern Recognition (CVPR)}, June 2018.

\bibitem{10.1145/3581783.3612242}
\BIBentryALTinterwordspacing
P.~Wang and X.~Yuan, ``Saunet: Spatial-attention unfolding network for image compressive sensing,'' in \emph{Proceedings of the 31st ACM International Conference on Multimedia}, ser. MM '23.\hskip 1em plus 0.5em minus 0.4em\relax New York, NY, USA: Association for Computing Machinery, 2023, p. 5099–5108. [Online]. Available: \url{https://doi.org/10.1145/3581783.3612242}
\BIBentrySTDinterwordspacing

\bibitem{isprs_vaihingen}
ISPRS, ``The isprs vaihingen dataset,'' \url{https://www.isprs.org/education/benchmarks/UrbanSemLab/default.aspx}, 2012, accessed: 2024-07-09.

\bibitem{9681903}
L.~Wang, R.~Li, C.~Duan, C.~Zhang, X.~Meng, and S.~Fang, ``A novel transformer based semantic segmentation scheme for fine-resolution remote sensing images,'' \emph{IEEE Geoscience and Remote Sensing Letters}, vol.~19, pp. 1--5, 2022.

\bibitem{landsat8}
U.~G. Survey, ``Landsat-8 dataset,'' \url{https://www.usgs.gov/landsat-missions/landsat-8}, 2013, accessed: 2024-07-09.

\bibitem{CBSD68}
T.~B.~S. Dataset and Benchmark, ``Cbsd68 dataset,'' \url{https://www2.eecs.berkeley.edu/Research/Projects/CS/vision/bsds/}, accessed: 2024-07-09.

\bibitem{kingma2014adam}
D.~P. Kingma and J.~Ba, ``Adam: A method for stochastic optimization,'' \emph{arXiv preprint arXiv:1412.6980}, 2014.

\bibitem{yolo}
\BIBentryALTinterwordspacing
G.~Jocher, ``{YOLOv5 by Ultralytics},'' May 2020. [Online]. Available: \url{https://github.com/ultralytics/yolov5}
\BIBentrySTDinterwordspacing

\bibitem{RN53}
\BIBentryALTinterwordspacing
l.~Peking Qing zhi yuan Shi~technology co., ``Metacam,'' 2024, accessed: 2024-04-10. [Online]. Available: \url{https://www.metacam.tech/}
\BIBentrySTDinterwordspacing

\bibitem{Hu:21}
\BIBentryALTinterwordspacing
C.~Hu, H.~Huang, M.~Chen, S.~Yang, and H.~Chen, ``Fouriercam: a camera for video spectrum acquisition in a single shot,'' \emph{Photon. Res.}, vol.~9, no.~5, pp. 701--713, May 2021. [Online]. Available: \url{https://opg.optica.org/prj/abstract.cfm?URI=prj-9-5-701}
\BIBentrySTDinterwordspacing

\bibitem{10.1063/5.0040424}
\BIBentryALTinterwordspacing
------, ``{Video object detection from one single image through opto-electronic neural network},'' \emph{APL Photonics}, vol.~6, no.~4, p. 046104, 04 2021. [Online]. Available: \url{https://doi.org/10.1063/5.0040424}
\BIBentrySTDinterwordspacing

\bibitem{10376842}
P.~Wang, L.~Wang, and X.~Yuan, ``Deep optics for video snapshot compressive imaging,'' in \emph{2023 IEEE/CVF International Conference on Computer Vision (ICCV)}, 2023, pp. 10\,612--10\,622.

\bibitem{opencv}
``Opencv library,'' \url{https://opencv.org/}, accessed: 2024-12-01.

\end{thebibliography}

\section{Biography Section}

\begin{IEEEbiography}
[{\includegraphics[width=1in,height=1.25in,clip,keepaspectratio]{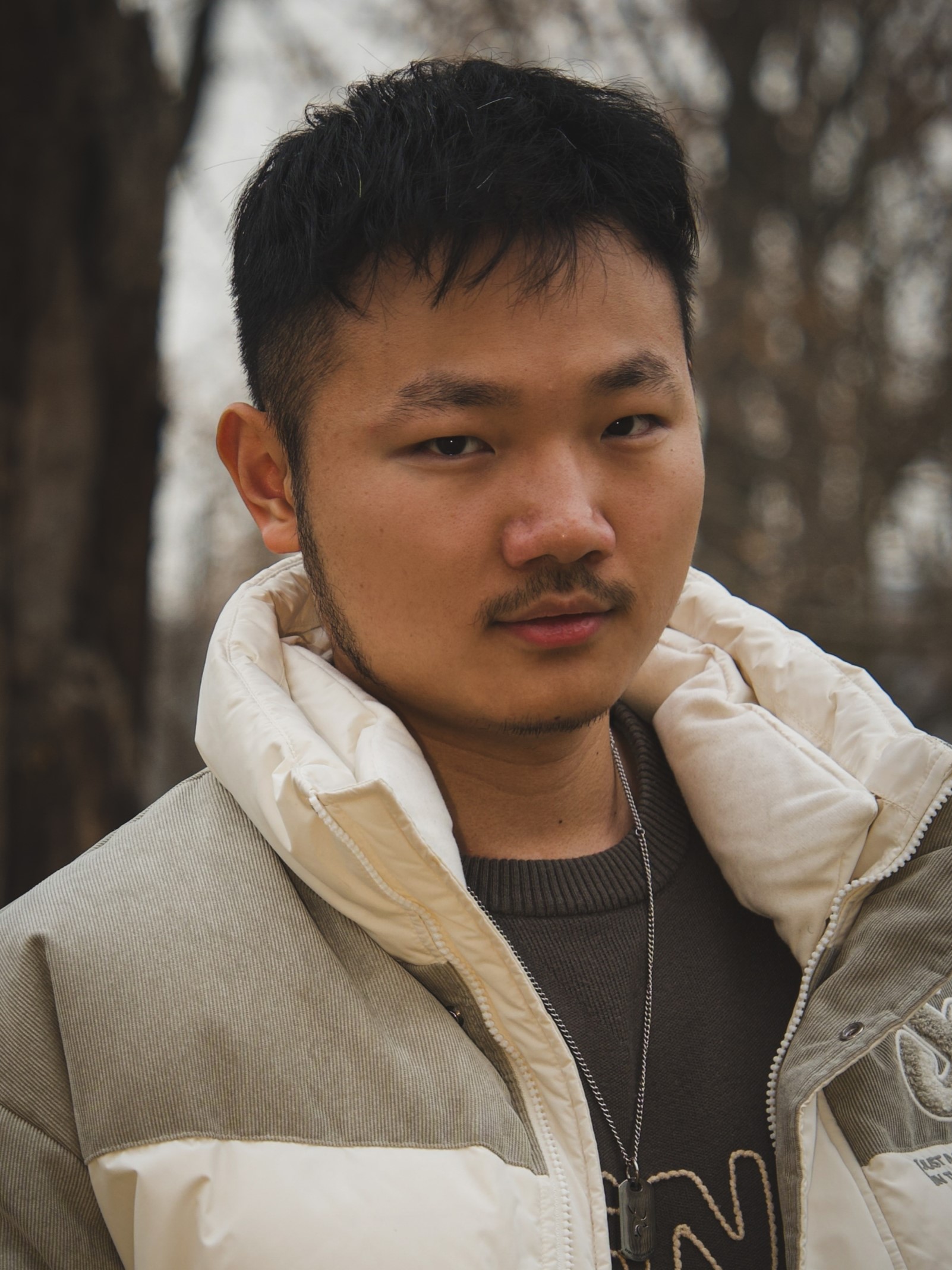}}]{Zhibin Wang}
Zhibin Wang received the B.E. degree from the School of Computer Science and Engineering, University of Electronic Science and Technology of China, Chengdu, China, in 2023, where he is currently pursuing the M.E. degree.

His research interests include compressed sensing and remote sensing image processing.
\end{IEEEbiography}

\begin{IEEEbiography}
[{\includegraphics[width=1in,height=1.25in,clip,keepaspectratio]{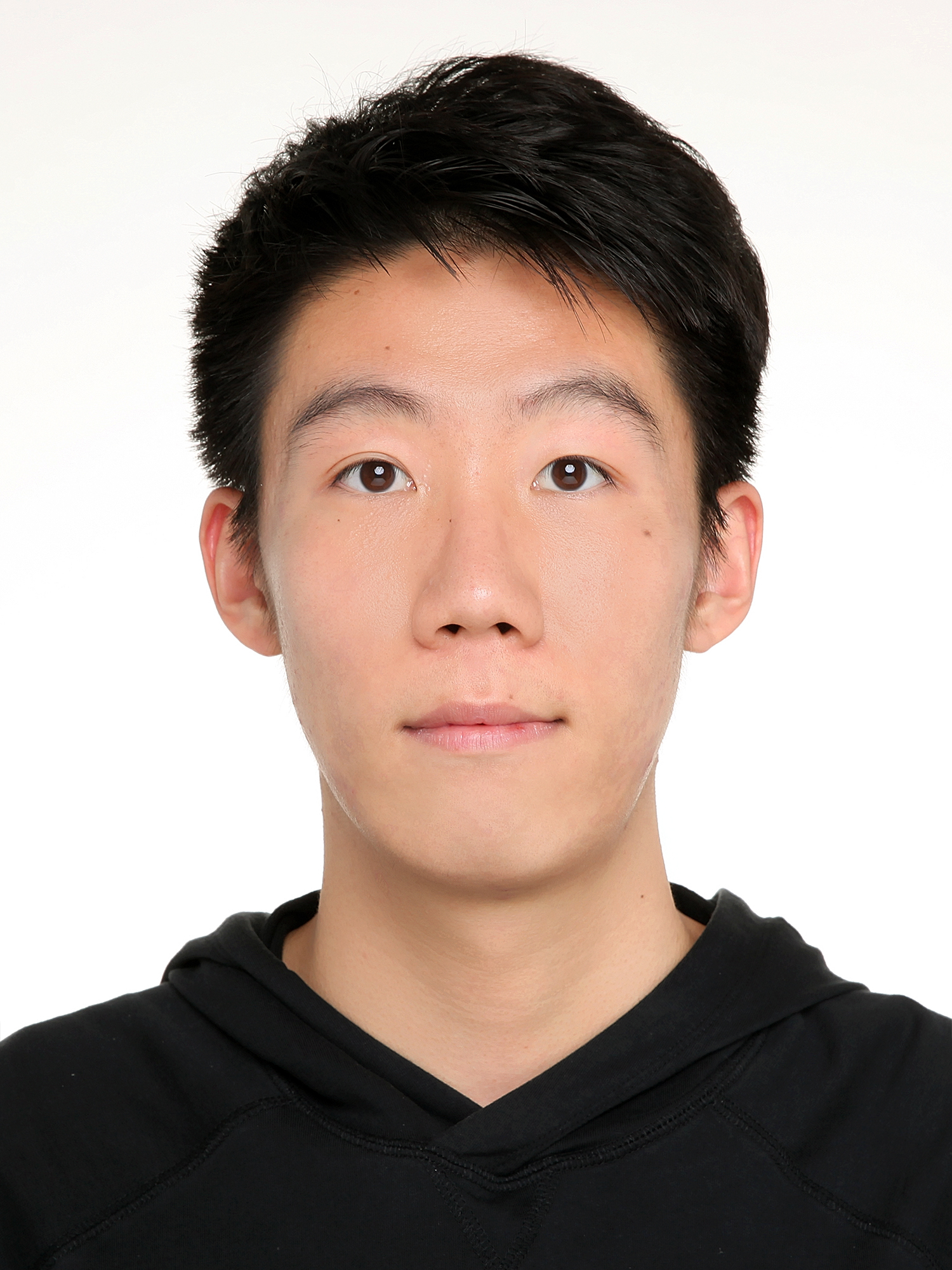}}]{Yanxin Cai}
Yanxin Cai received the B.E. degree in Control Engineering from Tongji University, Shanghai, in 2022. He is currently pursuing the M.E. degree with the Beijing Institute of Space Mechanics and Electricity, Beijing. 

His research interests include remote sensing image processing and intelligent optical remote sensing. 
\end{IEEEbiography}

\begin{IEEEbiography}
[{\includegraphics[width=1in,height=1.25in,clip,keepaspectratio]{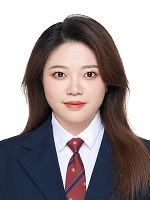}}]{Jiayi Zhou}
Jiayi Zhou received the B.E degree from the University of Electronic Science and Technology of China, Chengdu, China, in 2024. She is currently pursuing the M.E. degree at the same institution. 

Her research interests include computer vision and optical computational imaging.
\end{IEEEbiography}

\begin{IEEEbiography}
[{\includegraphics[width=1in,height=1.25in,clip,keepaspectratio]{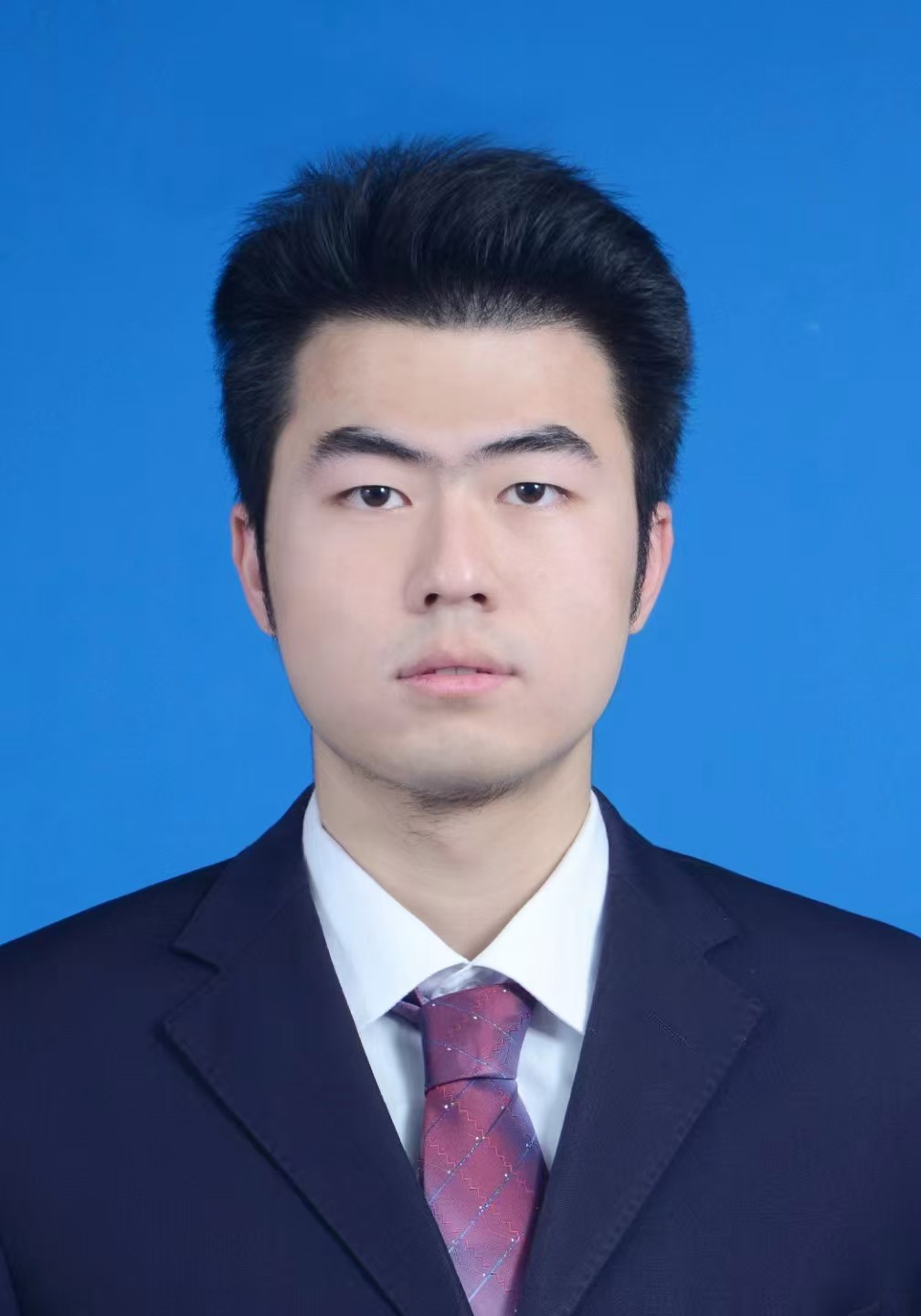}}]{Yangming Zhang}
Yangming Zhang received the B.E degree from the University of Electronic Science and Technology of China, Chengdu, China, in 2024. He is currently pursuing the Ph.D. degree at the University of Electronic Science and Technology of China. 

His research interests include image compression and decompression and optical imaging.
\end{IEEEbiography}

\begin{IEEEbiography}
[{\includegraphics[width=1in,height=1.25in,clip,keepaspectratio]{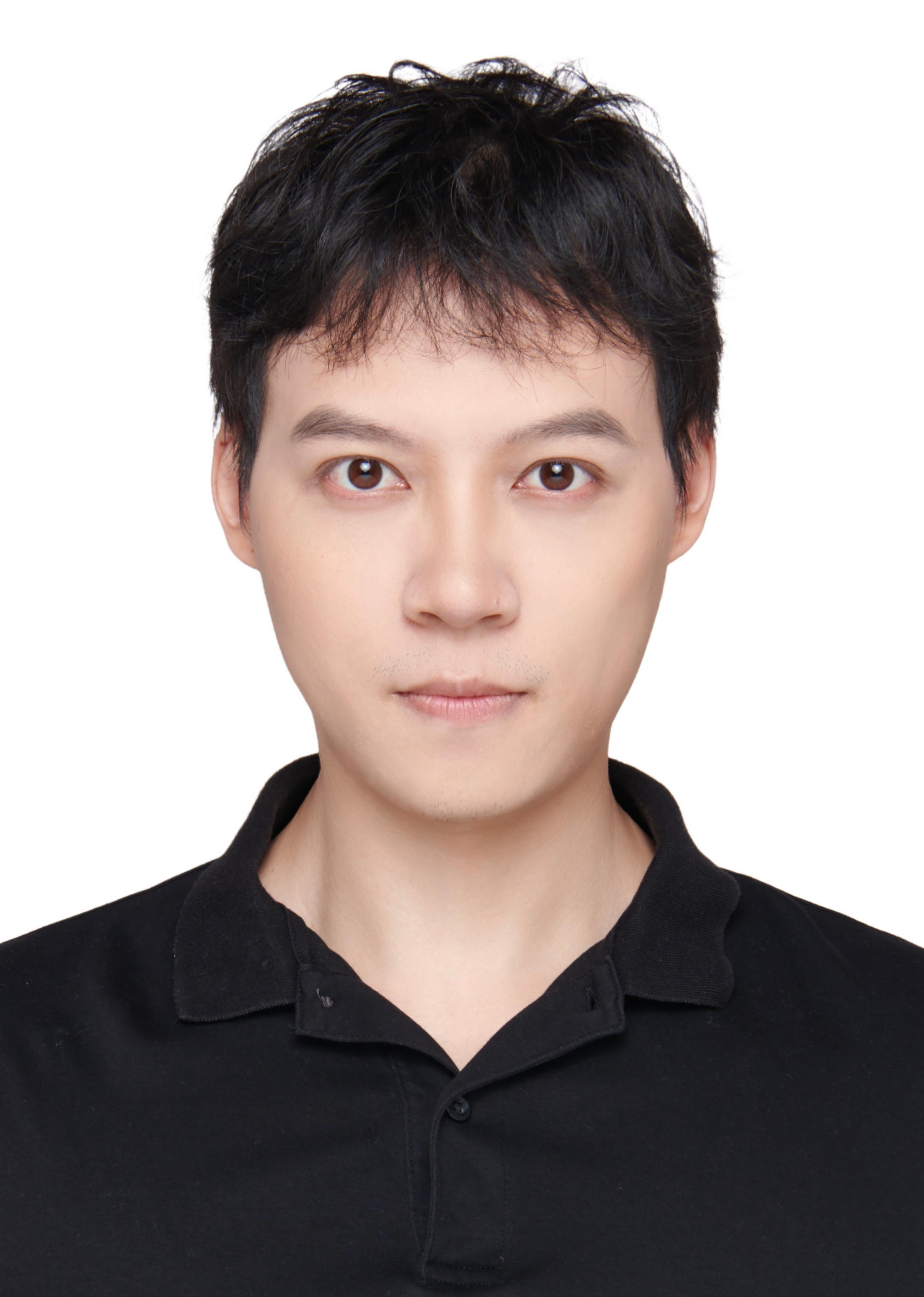}}]{Tianyu Li}
Tianyu Li received his Ph.D. degree from the School of Electrical and Computer Engineering at Purdue University, USA, in 2023. He is currently a postdoctoral researcher at the Center for Future Media, University of Electronic Science and Technology of China, located in Chengdu, China. 

His research focuses on image processing using statistical models and deep learning models, as well as time series forecasting and anomaly detection in time series.
\end{IEEEbiography}

\begin{IEEEbiography}
[{\includegraphics[width=1in,height=1.25in,clip,keepaspectratio]{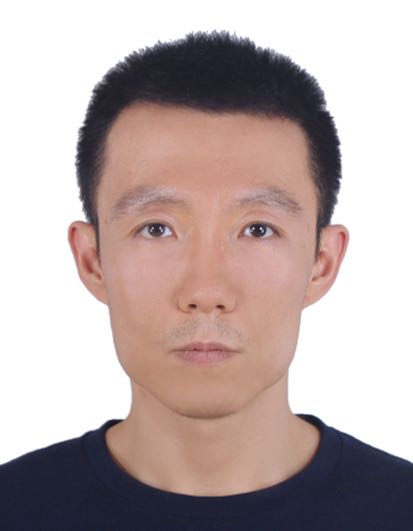}}]{Wei Li}
Wei Li received the Ph.D. degree in optics from the Beijing University of Technology, Beijing, China, in
2010. He is currently a Research Fellow with the Beijing Institute of Space Mechanics and Electricity, Beijing,
and a part-time Supervisor with the Beijing University of Technology. 

His main research fields are intelligent optical remote sensing, optical computing remote sensing, and advanced optical detection.
\end{IEEEbiography}

\begin{IEEEbiography}
[{\includegraphics[width=1in,height=1.25in,clip,keepaspectratio]{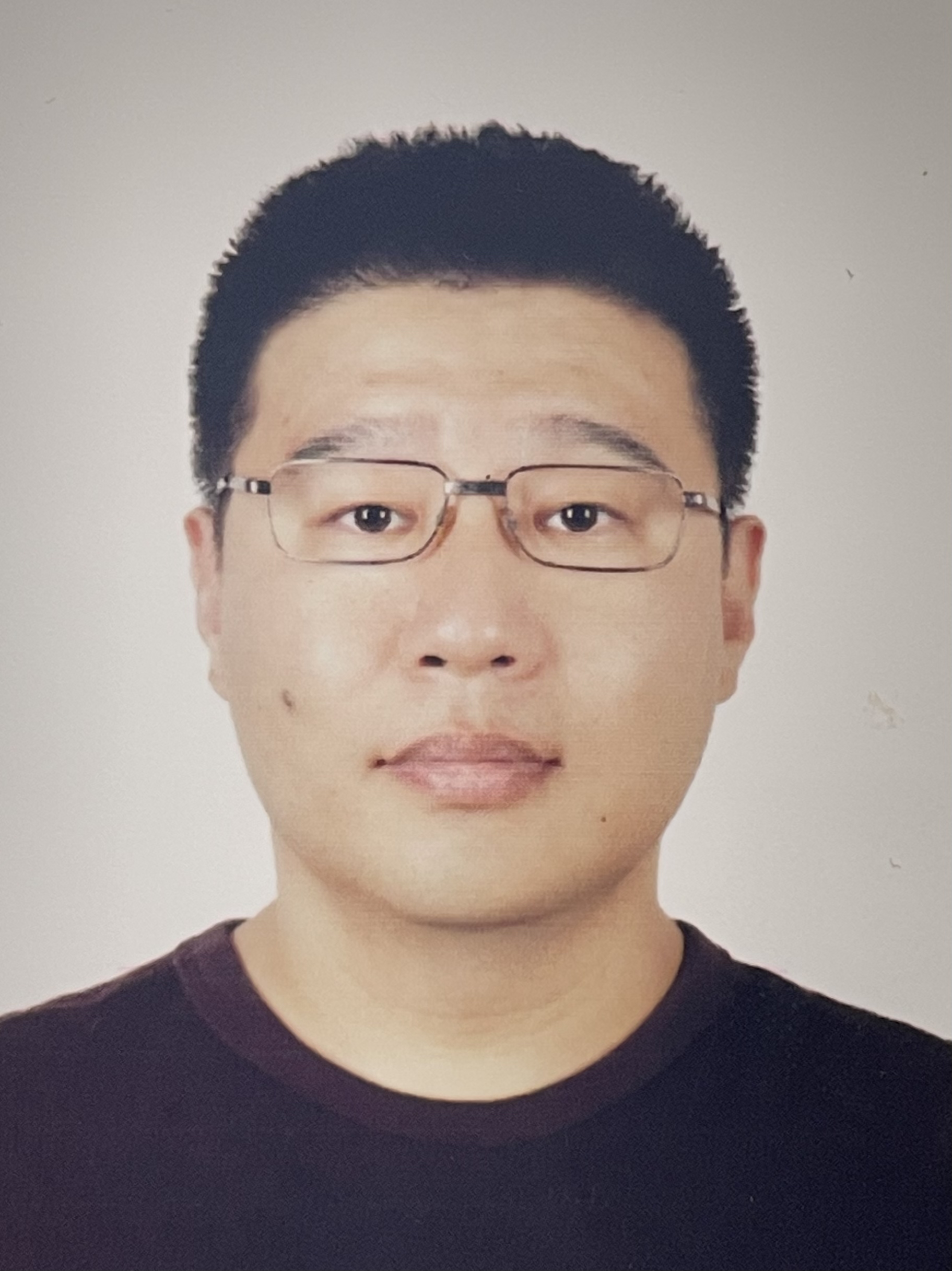}}]{Xun Liu}
Xun Liu received the Ph.D. degree in optics from the Institute of Physics, Chinese Academy of Sciences,
Beijing, China, in 2012. He is currently a Senior Engineer with the Beijing Institute of Space Mechanics and Electricity, Beijing. 

His research interests include intelligent optical remote sensing and advanced optical detection.
\end{IEEEbiography}

\begin{IEEEbiography}
[{\includegraphics[width=1in,height=1.25in,clip,keepaspectratio]{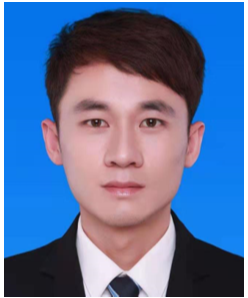}}]{Guoqing Wang}
Guoqing Wang (Member, IEEE) received the Ph.D. degree from the University of New South Wales, Sydney, NSW, Australia, in 2021. He is currently with the School of Computer Science and Engineering, University of Electronic of Science and Technology of China, Chengdu, China. He has authored and coauthored more than 40 scientific articles at top venues, including International Journal of Computer Vision, IEEE TRANSACTIONS
ON IMAGE PROCESSING, IEEE TRANSACTIONS ON INFORMATION FORENSICS AND SECURITY, International Conference on Computer Vision, and ACM Multimedia Conference. 

His research interests include machine learning and unmanned system, with special emphasis on cognition and embodied agents.
\end{IEEEbiography}

\begin{IEEEbiography}
[{\includegraphics[width=1in,height=1.25in,clip,keepaspectratio]{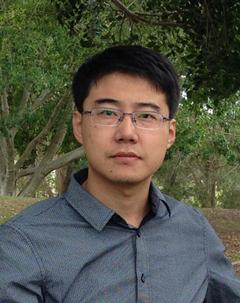}}]{Yang Yang}
Yang Yang (Senior Member, IEEE) received the bachelor’s degree from Jilin University, Changchun, China, in 2006, the master’s degree from Peking University, Beijing, China, in 2009, and the Ph.D. degree from The University of Queensland, Brisbane, QLD, Australia, in 2012, all in computer science. He is currently with the University of Electronic Science and Technology of China, Chengdu, China. 

His research interests include multimedia content analysis, computer vision, and social media analytics.
\end{IEEEbiography}

\vspace{11pt}

\vfill

\end{document}